\documentclass[a4paper,11pt]{article}
\pdfoutput=1 

\usepackage{jheppub} 

\usepackage{amsfonts,amscd,amssymb,amsmath,amsthm,latexsym}
\usepackage{verbatim}

\newcommand{\Natural}{\mathbb{N}}

\newcommand{\dx}{d^4x}

\newcommand{\nmu}{\nabla_{\mu}}
\newcommand{\nnu}{\nabla_{\nu}}

\newcommand{\FF}{\mathcal{F}}
\newcommand{\GG}{Q} 
\newcommand{\HH}{P} 
\newcommand{\GGP}{Q'}
\newcommand{\HHP}{P'}

\newtheorem{lemma}{Lemma}[section]

\title{\boldmath  Nonlocal de Sitter gravity and its exact cosmological  solutions   }

\author[a]{Ivan Dimitrijevic,}
\author[b,c,1]{Branko Dragovich,\note{Corresponding author.}}
\author[a]{Zoran Rakic}
\author[d]{and Jelena Stankovic}

\affiliation[a]{Faculty of Mathematics, University of Belgrade, \\ Studentski trg 16, 11158 Belgrade, Serbia}
\affiliation[b]{Institute of Physics, University of Belgrade,\\ Pregrevica 118, 11080 Belgrade, Serbia}
\affiliation[c]{Mathematical Institute, Serbian Academy of Sciences and Arts, \\ Kneza Mihaila 36, 11000 Belgrade, Serbia}
\affiliation[d]{Teacher Education Faculty, University of Belgrade, \\ Kraljice Natalije 43, 11000 Belgrade, Serbia}

\emailAdd{ivand@matf.bg.ac.rs}
\emailAdd{dragovich@ipb.ac.rs}
\emailAdd{zrakic@matf.bg.ac.rs}
\emailAdd{jelena.grujic@uf.bg.ac.rs}

\abstract{This paper is devoted to  a simple nonlocal de Sitter gravity model and its exact vacuum cosmological solutions. In the Einstein-Hilbert action with $\Lambda$ term, we introduce nonlocality by the following way: $R - 2 \Lambda = \sqrt{R-2\Lambda}\ \sqrt{R-2\Lambda} \to \sqrt{R-2\Lambda}\ F(\Box)\ \sqrt{R-2\Lambda} ,$ where ${F} (\Box) =  1 + \sum_{n= 1}^{+\infty} \big( f_n \Box^n + f_{-n} \Box^{-n} \big) $ is an analytic function of the d'Alembert-Beltrami operator $\Box$ and its inverse $\Box^{-1}$. By this  way,  $R$ and $\Lambda$ enter with the same form into nonlocal version as they are in the local one, and nonlocal operator $F(\Box)$ is dimensionless.  The corresponding equations of motion for gravitational field $g_{\mu\nu}$ are presented. The first step in finding some exact cosmological solutions is  solving the equation $\Box \sqrt{R-2\Lambda} =  q \sqrt{R-2\Lambda} , $ where $ q =\zeta \Lambda  \quad (\zeta \in \mathbb{R})$ is an eigenvalue and  $\sqrt{R-2\Lambda}$ is an eigenfunction of the operator $\Box .$  We presented and discussed several exact cosmological solutions for homogeneous and isotropic universe. One of these solutions mimics effects that are usually assigned to dark matter and dark energy. Some other solutions are examples of the nonsingular bounce ones in flat, closed and open universe. There are also singular and cyclic solutions. All these cosmological solutions are a result of nonlocality and do not exist in the local de Sitter case.}

\begin{document}
\maketitle
\flushbottom

\section{Introduction}

According to the Standard  Model of Cosmology (SMC), at the current cosmic time the universe approximately consists of 68 \% of  dark energy (DE), 27 \% of  dark matter (DM) and only  5 \% of  visible (ordinary) matter known from the Standard model of particle physics, see \cite{Planck2018}. It is worth noting  that  the SMC assumes  General Relativity (GR) as theory of the gravitational interaction, not only in the Solar system but also at the galactic and cosmological scales.  According to the SMC point of view, dark matter is responsible for observational dynamics inside galaxies and their clasters, while dark    energy acts as repulsive force and causes accelerated expansion of the universe. The SMC is also known as the $\Lambda$CDM model, what means that DE corresponds to the cosmological constant and that DM is in a cold state. However, despite many efforts to confirm existence of DM and DE either in the sky or  in the laboratory experiments, they are still not discovered and remain hypothetical.

There is a common opinion that general relativity is one of the most beautiful and successful physical theories \cite{ellis}. From phenomenological point of view, GR has had several significant  confirmed  predictions. However, GR  as a theory of gravitation has not been verified at the galactic and cosmological scales. Hence, in spite of remarkable successes, it is reasonable to doubt in its validity in description and understanding of all  astrophysical and cosmological gravitational phenomena. Moreover, GR solutions for the black holes as well as for the beginning of the universe contain singularity and it
means that GR should be modified in the vicinity of these singularities, see,e.g. \cite{novello,koshelev}. It is also well known that GR  is nonrenormalizable theory from the quantum point of view. Let us also mention that any other physical theory has its domain of validity which is usually  constrained by spacetime scale, complexity of the system under consideration, or by some parameters. There is no a priory reason that GR should be appropriate at all spacetime scales. Keeping all this in mind, it follows that general relativity is not a final theory of gravitation and that its extension is desirable.

Since  no new physical principle is presently known that could tell us in what direction to look for theoretical generalization of GR, there are many approaches to its modification, see \cite{faraoni,nojiri,clifton,nojiri1,capozziello,dragovich0} as some reviews. Despite many attempts, there is not yet generally accepted modification of general relativity. One of the current and attractive approaches is nonlocal modified gravity, see, e.g. \cite{biswas1,biswas2,biswas3,biswas4,biswas5,deser}. The idea behind nonlocality is that dynamics of the gravitational field may depend not only on its first and second spacetime derivatives but also on all higher derivatives.  Practically, it means that the Einstein-Hilbert action  should be modified
by an additional term that contains higher order degrees of d'Alembert-Beltrami operator  $\Box = \nabla_\mu \nabla^\mu = \frac{1}{\sqrt{-g}} \partial_\mu (\sqrt{-g} g^{\mu\nu}\partial_\nu) .$ So far $\Box$ has been mainly used in one of the following two forms: (i) non-polynomial analytic expansion $F(\Box) = \sum_{n=0}^{+\infty} f_n \ \Box^n ,$ see \cite{koshelev1,koshelev2,koshelev3,koshelev4,eliz,koivisto,capozziello2009,craps} as some examples  or (ii)  some  inverse manner $ \Box^{-1} ,$ see, e. g. \cite{deser,woodard,maggiore,barvinsky2012,maggiore1,golovnev}.

A part of motivation to use nonlocal operator in the form  $F(\Box) = \sum_{n=0}^{+\infty} f_n \ \Box^n $ comes from string field theory
\cite{arefeva1}  and $p$-adic string theory \cite{dragovich1}. Also this type of nonlocality improves quantum renormalizability \cite{stelle,modesto1,modesto2}.

Some of the nonlocal gravity models that have been so far considered with analytic nonlocality are particular cases of the action
\begin{align} \label{eq1.1}
S = \frac{1}{16 \pi G} \int d^4 x \ \sqrt{-g}\ \big(R- 2 \Lambda  + P(R){\FF}(\Box) Q(R)\big) ,
\end{align}
where $\Lambda$ is the cosmological constant, $P(R)$ and $Q(R)$ are some differentiable functions of the Ricci scalar $R$, see \cite{dimitrijevic1,dimitrijevic2,dimitrijevic3,dimitrijevic4,dimitrijevic5,dimitrijevic6,dimitrijevic7,dimitrijevic8,dimitrijevic9,dimitrijevic10,
dimitrijevic11,dimitrijevic12} and references therein. The case $P(R) = Q(R) =R$ and $\Lambda =0$
has attracted a lot of attention, i.e. nonlocal $R^2$ gravity, see \cite{biswas1,koshelev2} and references therein.

A very special case of \eqref{eq1.1} is  
\begin{align} \label{eq1.2a}
S &= \frac{1}{16 \pi G} \int d^4 x \ \sqrt{-g}\ \big(R - 2 \Lambda + \sqrt{R - 2\Lambda} \mathcal{F}(\Box) \sqrt{R - 2\Lambda} \big), \\
  &= \frac{1}{16 \pi G} \int d^4 x \ \sqrt{-g}\ \sqrt{R - 2\Lambda} F(\Box) \sqrt{R - 2\Lambda}
\label{eq1.2b}
\end{align}
where $F(\Box) = 1 + \mathcal{F}(\Box) = 1 + \sum_{n=1}^{+\infty} f_n \Box^n ,$ see \cite{dimitrijevic10}. In fact, in the Einstein-Hilbert action  a nonlocal
term is introduced so that dependence on $R$ and $\Lambda$ remains the same as in the local $(F(\Box) = 1)$ case. By this way, nonlocal model \eqref{eq1.2a}
is introduced by minimal change of the  Einstein-Hilbert action  with nonlocal dimensionless operator ($\mathcal{F}(\Box) =  \sum_{n=1}^{+\infty} f_n \Box^n $).

Importance of  model \eqref{eq1.2a} is not only in its simple form but also in its cosmological solutions in flat space-time: (i) $a_1(t) = A t^{\frac{2}{3}} e^{\frac{\Lambda}{14} t^2} $ and (ii) $a_2(t)= A e^{\frac{\Lambda}{6} t^2} .$
Solution $a_1(t)$  mimics interplay of dark matter and dark energy in a good agreement with  cosmological observations \cite{Planck2018,dimitrijevic10}. Solution $a_2 (t)$ is the nonsingular bounce one.  The corresponding explicit expressions of the nonlocal operator are:  $\mathcal{F}(\Box_1) = \frac{7}{3 e} \frac{\Box_1}{\Lambda}\ \exp{\big(-\frac{7}{3}\frac{\Box_1}{\Lambda}\big)}$ and $\mathcal{F}(\Box_2) = \frac{\Box_2}{ e \Lambda}\ \exp{\big(-\frac{\Box_2}{\Lambda}\big)} ,$ respectively, see also Section \ref{Sec.2}.
Note that $\frac{\Box}{\Lambda}$ is dimensionless operator and no new  mass (energy) parameter is introduced  in this model.

Another interesting example of \eqref{eq1.1} gets when $P(R) = Q(R) = R - 4\Lambda$, which can be derived from \eqref{eq1.2a} under approximation
$|R|\ll |2 \Lambda|$ \cite{dimitrijevic11,dimitrijevic12}. In the paper \cite{dimitrijevic11} two cosmological solutions are presented: (i) $a(t) = A \sqrt{t} e^{\frac{\Lambda}{4}t^2} ,$ which mimics properties similar to an interference between radiation and the dark energy, and (ii) nonsingular bounce solution $a(t) =A e^{\Lambda t^2} .$ The paper \cite{dimitrijevic12} contains cosmological solutions which demonstrate how this kind of gravitational nonlocality can change background topology.

In this paper we consider model \eqref{eq1.2a} with extension of  nonlocal operator  $\FF(\Box) =  \sum_{n=1}^{+\infty} f_n \Box^n $ to
$\FF(\Box) =   \sum_{n=1}^{+\infty} \big( f_n \Box^n  +  f_{-n} \Box^{-n} \big) $ and finding som new cosmological solutions.

Structure of the paper is as follows. In Section \ref{Sec.2},  we introduce a simple nonlocal de Sitter gravity model giving its action and present the corresponding equations of motion. We also consider the suitable form of nonlocal operator $\mathcal{F} (\Box)$ as an analytic function with respect to $\Box$ and $\Box^{-1} .$ Presentation of two known solutions and finding some new exact vacuum cosmological solutions is subject of Section \ref{Sec.3}. All obtained cosmological solutions of this nonlocal de Sitter gravity are discussed in Section \ref{Sec.4}. Some concluding remarks,
with list of main results and plan for future works, are contained in Section \ref{Sec.5}. At the end, there is also an appendix related to the derivation of equations of motion.

 \section{Nonlocal de Sitter gravity}
\label{Sec.2}

\subsection{Action}

Our nonlocal gravity model is given by its action

\begin{align} \label{eq2.1}
S = \frac{1}{16 \pi G} \int d^4 x \ \sqrt{-g}\ \sqrt{R - 2\Lambda}\ F(\Box)\ \sqrt{R - 2\Lambda} ,
\end{align}
where $F (\Box)$ is the following formal expansion in terms of the d'Alembert-Beltrami operator $\Box$:
\begin{align}  \label{eq2.2}
F(\Box) = 1+ \FF (\Box) = 1 + \FF_{+} (\Box) + \FF_{-} (\Box) , \quad \FF_{+} (\Box) =\sum_{n=1}^{+\infty} f_n \ \Box^n , \ \FF_{-} (\Box) =\sum_{n= 1}^{+\infty} f_{-n} \ \Box^{-n} .
\end{align}

 When $F(\Box)= 1$, i.e. $\FF (\Box) = 0$,
then  model \eqref{eq2.1} becomes
local and coincides with Einstein-Hilbert action with cosmological constant $\Lambda$:
\begin{align} \label{eq2.3}
  S_0 = \frac{1}{16 \pi G} \int d^4 x \ \sqrt{-g}\ \sqrt{R - 2\Lambda} \ \sqrt{R - 2\Lambda} = \frac{1}{16 \pi G} \int d^4 x \ \sqrt{-g}\ (R - 2 \Lambda) .
\end{align}
 It is worth pointing out that action \eqref{eq2.1} can be obtained in a very simple and natural way from action \eqref{eq2.3}
 by embedding nonlocal operator \eqref{eq2.2} in symmetric product form of $R- 2\Lambda$, that is $\sqrt{R - 2 \Lambda}\ \sqrt{R - 2 \Lambda} .$
Action \eqref{eq2.1} does not contain matter term and this is intentionally done to better see possible role of this nonlocal model  in effects
usually assigned to dark matter and dark energy.

At the beginning, let the nonlocal operator $\FF (\Box)$ be in general form  \eqref{eq2.2}, that is in all powers of $\Box$ and $\Box^{-1}$ with unknown real coefficients $f_n$ and $f_{-n}$, where $ 1 \leq n < +\infty$. These coefficients $f_n$ and $f_{-n}$ will be  presented later in the simple form which satisfies necessary conditions for exact cosmological solutions.

\subsection{Equations of motion}

The next step is obtaining the equations of motion (EoM) for model \eqref{eq2.1}. To this end, we will start with  more general case, that is
\begin{align} \label{eq2.4}
S = \frac{1}{16 \pi G} \int d^4 x \ \sqrt{-g}\ \big(R- 2 \Lambda +  P(R)\ \FF(\Box)\ Q(R) \big) ,
\end{align}
 where $P (R)$ and $Q(R)$ are some differentiable functions of $R$, while $\FF(\Box) = \sum_{n=1}^{+\infty} f_n \ \Box^n + \sum_{n=1}^{+\infty} f_{-n} \ \Box^{-n}$ remains the same as in \eqref{eq2.2}. A short version of derivation of
 EoM for model \eqref{eq2.4} is presented in  Appendix,  a similar case can be seen in \cite{dimitrijevic9}.
 According to expressions (A.23) and (A.24) in the Appendix, equations of motion for nonlocal model \eqref{eq2.4} are
 \begin{align} \label{eq2.5}
 \hat G_{\mu\nu} = G_{\mu\nu} +\Lambda g_{\mu\nu} - \frac 12 g_{\mu\nu} \HH\; \FF(\Box)\GG + R_{\mu\nu}W -K_{\mu\nu}W + \frac 12 \Omega_{\mu\nu} = 0,
 \end{align}
 where
 \begin{align}
 &W =  P'(R)\ \FF (\Box)\ Q(R) + Q'(R)\ \FF (\Box) P(R) , \quad K_{\mu\nu} = \nabla_\mu \nabla_\nu - g_{\mu\nu}\Box , \label{eq2.6a} \\
  &\Omega_{\mu\nu} =  \sum_{n=1}^{+\infty} f_n \sum_{\ell=0}^{n-1} S_{\mu\nu}(\Box^\ell \HH, \Box^{n-1-\ell} \GG)  -\sum_{n=1}^{+\infty} f_{-n} \sum_{\ell=0}^{n-1} S_{\mu\nu}(\Box^{-(\ell+1)} \HH, \Box^{-(n-\ell)} \GG), \label{eq2.6b} \\
  &S_{\mu\nu} (A, B) = g_{\mu\nu} \big(\nabla^\alpha A \ \nabla_\alpha B + A \Box B \big) - 2\nabla_\mu A\ \nabla_\nu B . \label{eq2.6c}
\end{align}
and $P'\ (Q')$ means derivative of $P\ (Q)$ with respect to scalar curvature $R$.

Equations of motion \eqref{eq2.5} look very complicated comparing to their local (Einstein) counterpart $G_{\mu\nu} +\Lambda g_{\mu\nu} = 0$.
Finding some solutions of  \eqref{eq2.5} is a difficult task. Nevertheless, as we will see, one can find some exact cosmological solutions in the  case  when
$P =Q = \sqrt{R- 2 \Lambda}$.

Now, let us first consider the case $Q(R) = P(R)$. The corresponding EoM are
\begin{align} \label{eq2.7a}
  &G_{\mu\nu}+ \Lambda g_{\mu\nu} - \frac{g_{\mu\nu}}{2} P(R) \FF (\Box) P(R) + R_{\mu\nu} W - K_{\mu\nu} W + \frac 12 \Omega_{\mu\nu} = 0 , \\
&W = 2 P'(R)\ \FF (\Box)\ P(R), \quad K_{\mu\nu} = \nabla_\mu \nabla_\nu - g_{\mu\nu}\Box , \label{eq2.7b} \\
&\Omega_{\mu\nu} =  \sum_{n=1}^{+\infty} f_n \sum_{\ell=0}^{n-1} S_{\mu\nu}(\Box^\ell P, \Box^{n-1-\ell} P)  -\sum_{n=1}^{+\infty} f_{-n} \sum_{\ell=0}^{n-1} S_{\mu\nu}(\Box^{-(\ell+1)} P, \Box^{-(n-\ell)} P). \label{eq2.7c}
\end{align}

If $P(R)$ is an eigenfunction of the corresponding d'Alembert-Beltrami operator $\Box$, and consequently also of its inverse $\Box^{-1}$, i.e.  holds
\begin{align} \label{eq2.8}
\Box P(R) = q  P (R), \quad \Box^{-1} P(R) = q^{-1} P(R) ,  \quad \FF (\Box) P(R) = \FF (q) P(R) , \quad q\neq 0 ,
\end{align}
 then
\begin{align}
  &W = 2 \FF(q) P' P,  \qquad \FF(q) = \sum_{n=1}^{+\infty} f_n \ q^n + \sum_{n=1}^{+\infty} f_{-n} \ q^{-n} , \label{eq2.8} \\
 &S_{\mu\nu}(\Box^{\ell} P, \Box^{(n- 1 -\ell)} P)  = q^{n-1} S_{\mu\nu} (P, P),  \label{eq2.8a} \\
 &S_{\mu\nu}(\Box^{-(\ell+1)} P, \Box^{-(n-\ell)} P) = q^{-n-1} S_{\mu\nu} (P, P),  \label{eq2.8b} \\
 &S_{\mu\nu}(P, P) = g_{\mu\nu} \big(\nabla^\alpha P \ \nabla_\alpha P + P \Box P \big) - 2\nabla_\mu P\ \nabla_\nu P,  \label{eq2.8b}\\
 &\Omega_{\mu\nu} = \FF'(q) S_{\mu\nu}(P,P), \qquad \FF'(q) = \sum_{n=1}^{+\infty} n\ f_n \ q^{n-1} -
 \sum_{n=1}^{+\infty} n\ f_{-n} \ q^{-n-1} , \label{eq2.9}
\end{align}
and
\begin{equation} \label{eq2.10}
  G_{\mu\nu}+ \Lambda g_{\mu\nu} - \frac{g_{\mu\nu}}{2}  \FF (q)P^2 + 2 \FF(q) R_{\mu\nu} P P' - 2 \FF(q) K_{\mu\nu} P P' + \frac 12 \FF'(q) S_{\mu\nu}(P,P) = 0.
\end{equation}
The last equation transforms to
\begin{align} \nonumber
  &\left(G_{\mu\nu}+ \Lambda g_{\mu\nu}\right)\left(1 + 2 \FF(q) P P'\right) + \FF(q)g_{\mu\nu}\left(-\frac 12 P^2 + P P'(R-2\Lambda)\right) \\
 &- 2 \FF(q) K_{\mu\nu} P P' + \frac 12 \FF'(q) S_{\mu\nu}(P,P) = 0. \label{eq2.11}
\end{align}

Let now $P= \sqrt{R-2\Lambda}$, then  $PP' = \frac 12 $ and
\begin{equation}
\Box \sqrt{R - 2\Lambda} = q  \sqrt{R - 2\Lambda} = \zeta \Lambda  \sqrt{R - 2\Lambda} , \quad \zeta\Lambda \neq 0 , \label{eq2.11a}
\end{equation}
where $q = \zeta \Lambda$ and $q^{-1}= \zeta^{-1} \Lambda^{-1}$ \ ($\zeta$ -- dimensionless) follows from dimensionality of equalities \eqref{eq2.11a}.
Since $P= \sqrt{R-2\Lambda}$,  EoM \eqref{eq2.11} simplify to
\begin{equation} \label{eq2.12}
  \left(G_{\mu\nu}+ \Lambda g_{\mu\nu}\right)\left(1 + \FF(q)\right) + \frac 12 \FF'(q) S_{\mu\nu}(\sqrt{R-2\Lambda},\sqrt{R-2\Lambda}) = 0.
\end{equation}
It is evident that EoM \eqref{eq2.12} are satisfied if
\begin{align} \label{eq2.13}
\mathcal{F} (q) = -1  \quad \text{and} \quad  \mathcal{F}' (q) = 0 .
\end{align}
In this case, nonlocal action \eqref{eq2.1} has null value.
Nonlocal operator $\mathcal{F} (\Box)$, that satisfies conditions \eqref{eq2.13} in model \eqref{eq2.1},
can be taken in the symmetric form
\begin{equation}
F(\Box) = 1 +  \mathcal{F} (\Box), \quad \mathcal{F} (\Box) = \sum_{n=1}^{+\infty} \tilde{f}_n \Big[\Big(\frac{\Box}{q}\Big)^n  + \Big(\frac{q}{\Box}\Big)^n \Big] , \label{eq2.13a}
\end{equation}
where $\tilde{f}_n$ are now dimensionless coefficients.
It is easy to prove that
 $\mathcal{F}(\Box)$  presented in the following simple symmetric form:
\begin{align} \label{eq2.14}
 \mathcal{F}(\Box) = \sum_{n=1}^{+\infty} \tilde{f}_n \Big[\Big(\frac{\Box}{q}\Big)^n  + \Big(\frac{q}{\Box}\Big)^n \Big] = - \frac{1}{2 e} \Big(\frac{\Box}{q} e^{\frac{\Box}{q}} + \frac{q}{\Box} e^{\frac{q}{\Box}}\Big) , \quad q \neq 0 ,
\end{align}
satisfies conditions \eqref{eq2.13}. The symmetry is with respect to the transformation $\Big(\frac{\Box}{q}\Big) \leftrightarrow \Big(\frac{q}{\Box}\Big) .$
Operator $\FF (\Box)$, defined in \eqref{eq2.14}, has $\sqrt{R - 2\Lambda}$  as its eigenfunction with  eigenvalue $-1$, that is
\begin{align} \label{eq2.15}
- \frac{1}{2 e} \Big(\frac{\Box}{q} e^{\frac{\Box}{q}}
+ \frac{q}{\Box} e^{\frac{q}{\Box}}\Big) \sqrt{R - 2\Lambda} = - \sqrt{R - 2\Lambda} \quad \text{whenever}  \quad \Box \sqrt{R - 2\Lambda} = q \sqrt{R - 2\Lambda}.
\end{align}

Since the d'Alembert-Beltrami operator $\Box$ and Ricci scalar $R$ depend on spacetime metric, it is  essential to find suitable metric that will
satisfy \eqref{eq2.15}.
As we will see in the next section, there are several  exact cosmological solutions that realize \eqref{eq2.15}  and eigenvalue $q$ is proportional to $\Lambda$, i.e. $q = \zeta \Lambda$, where $\zeta \neq 0$ is a definite dimensionless constant for each concrete case. Moreover,
it has to be $q = \zeta \Lambda$, since there is no other parameter than $\Lambda$ which is of the same  dimension as $\Box$  in this nonlocal
gravity model. Hence, nonlocal operator \eqref{eq2.14} can be rewritten as
\begin{align} \label{eq2.16}
 \mathcal{F} (\Box) = \sum_{n=1}^{+\infty} \tilde{f}_n \Big[\Big(\frac{\Box}{\zeta\Lambda}\Big)^n  + \Big(\frac{\zeta\Lambda}{\Box}\Big)^n \Big] = - \frac{1}{2 e} \Big(\frac{\Box}{\zeta \Lambda} e^{\frac{\Box}{\zeta \Lambda}} + \frac{\zeta \Lambda}{\Box} e^{\frac{\zeta \Lambda}{\Box}}\Big) , \quad \zeta\Lambda \neq 0,
 \end{align}
where for some specific $\Box$ holds  $\Box \sqrt{R - 2\Lambda} = \zeta \Lambda \sqrt{R - 2\Lambda} .$

Note that representation of $\mathcal{F} (\Box)$ by exponential function in the form \eqref{eq2.16} is not unique and can be written in the following more general form
\begin{align} \label{eq2.17}
 \mathcal{F} (\Box) = - \frac{1}{2} e^{(\mp 1)}\ \Big(\frac{\Box}{\zeta \Lambda}\ e^{\big(\pm\frac{\Box}{\zeta \Lambda}\big)} + \frac{\zeta \Lambda}{\Box}\ e^{\big(\pm\frac{\zeta \Lambda}{\Box}\big)}\Big).
 \end{align}

\section{Cosmological solutions}
\label{Sec.3}

We are now interested in some exact cosmological  solutions of  nonlocal gravity model \eqref{eq2.1}
with equations of motion \eqref{eq2.12}.

At the cosmological scale the universe is homogeneous and isotropic with
the Friedmann-Lema\^{i}tre-Robertson-Walker (FLRW) metric
\begin{align} \label{eq3.1}
ds^2  = -dt^2 + a^2(t) \Big(\frac{dr^2}{1- k r^2} + r^2\ d\theta^2 + r^2 \sin^2{\theta} \ d\varphi^2 \Big) , \quad (c =1) . \ \ k = 0, \pm 1 ,
\end{align}
where $a(t)$ is the  cosmic scale factor.
Our primary interest is to find cosmological scale factor $a(t)$ that satisfies equations \eqref{eq2.11a}, i.e.
\begin{align}  \label{eq3.2}
  \Box\ \sqrt{R - 2 \Lambda} = q \ \sqrt{R - 2 \Lambda} ,\quad \Box^{-1}\ \sqrt{R - 2 \Lambda} = q^{-1} \ \sqrt{R - 2 \Lambda} , \quad q =\zeta\Lambda \neq 0 ,
\end{align}
where $\Box$, $R$ and the Hubble parameter $H$ in the FLRW universe depend on $a(t)$ as follows:
\begin{align}  \label{eq3.3}
&\Box = -\frac{\partial^2}{\partial t^2} - 3 H(t)\ \frac{\partial}{\partial t} , \quad H (t) = \frac{\dot{a}}{a} \\
&R(t) = 6 \Big(\frac{\ddot{a}}{a} + \big(\frac{\dot{a}}{a}\big)^2 + \frac{k}{a^2}\Big) , \quad \dot{a} = \frac{\partial a}{\partial t} . \label{eq3.4}
\end{align}
If $a(t)$ is a solution of equation \eqref{eq3.2}, then it is also solution of equations of motion \eqref{eq2.15} with the corresponding two conditions \eqref{eq2.13} on nonlocal operator: $\mathcal{F}(q) = -1 $ and $\mathcal{F}'(q) = 0.$

Before to start concrete investigation of the possible time dependent scale factors $a(t)$, let us note that there is not  the Minkowski space solution. It follows from the fact that one can not take $\Lambda = 0$ in nonlocal operator $\mathcal{F}(\Box)$, see \eqref{eq2.16} and \eqref{eq3.2}. According to \eqref{eq3.2},
there is not also (anti-)de Sitter solution, since in (anti-)de Sitter case $\zeta = 0$ and consequently $q = 0.$  However, if there is the absence of terms with $\Box^{-1} ,$ then there exist both Minkowski and (anti-)de Sitter solutions. Namely, in the  Minkowski case $\Lambda = R = k = 0$ and EoM \eqref{eq2.12} are satisfied since $\mathcal{F} (0) = 0$ and $\mathcal{F}' (0) = 0 .$

\subsection{Cosmological solutions in the flat universe \ ($k = 0$)}

\subsubsection{Solutions  of the form $a(t) = A\ t^n\ e^{\gamma t^2}, \  (k= 0)$ }
\label{3.2.1}

According to our previous paper \cite{dimitrijevic10}, there are two solutions of the form
\begin{align}
a(t) = A\ t^n\ e^{\gamma t^2} ,  \quad k= 0 ,
\end{align}
where $ n$ and $\gamma$ are some definite real constants. The corresponding $\Box, H(t)$ and $R(t)$ are:
\begin{align} \label{eq3.3a}
&\Box = -\frac{\partial^2}{\partial t^2} - 3 \big(n t^{-1} + 2\gamma t\big)\ \frac{\partial}{\partial t}, \quad  H(t) = n t^{-1} + 2\gamma t , \\
&R(t) = 6 \big(-n t^{-2} + 2\gamma  + 2 \big(n t^{-1} + 2\gamma t\big)^2 \big) .  \label{eq3.3b}
\end{align}

Inserting \eqref{eq3.3a} and \eqref{eq3.3b} in \eqref{eq3.2} one obtains a system of six equations:
\begin{align*}
&n^2(2-3n)(2n-1)^2 = 0 ,\\
&n(2n-1)(-nq + 2n^2 q -\Lambda +n\Lambda + 6 \gamma + 24n \gamma - 36 n^2 \gamma) = 0 , \\
&n(2n-1) (-q \Lambda + 6q \gamma + 24 n q \gamma + 3 \Lambda \gamma + 54 \gamma^2 - 72 n \gamma^2) = 0, \\
&q \Lambda^2 - 12q \Lambda\gamma - 48nq \Lambda\gamma + 36q \gamma^2 + 144nq \gamma^2 + 864n^2 q \gamma^2 -24\Lambda \gamma^2 \\
&\qquad  - 72 n \Lambda\gamma^2 + 144 \gamma^3+ 1008 n \gamma^3 + 1728 n^2 \gamma^3 = 0 , \\
&\gamma^2 (-q \Lambda + 6 q \gamma + 24n q \gamma - 3\Lambda \gamma + 18 \gamma^2  + 108 n\gamma^2) = 0, \\
&\gamma^4 (q + 6 \gamma) = 0 .
\end{align*}
The above system of equations is satisfied in the following two cases:
\begin{align}
1.& \quad n =\frac{2}{3} , \ \ \gamma = \frac{\Lambda}{14} , \ \  q = - \frac{3}{7}\Lambda , \label{eq3.3c} \\
2.& \quad  n=0 , \quad \gamma = \frac{1}{6}\Lambda , \ \ q = -\Lambda .  \label{eq3.3d}
\end{align}

According to \eqref{eq3.3c} and \eqref{eq3.3d}, there are two solutions in flat space:
\begin{align}
 &a_1(t)  = A\ t^{\frac{2}{3}}\ e^{\frac{\Lambda}{14} t^2} , \quad k =0, \quad \mathcal{F}(-\frac{3}{7} \Lambda) = -1 , \ \ \mathcal{F}'(-\frac{3}{7} \Lambda) = 0 ,  \label{eq3.4a} \\
 &a_2(t)  = A\  e^{\frac{\Lambda}{6} t^2}, \quad k =0, \quad \mathcal{F}(-\Lambda) = -1 , \ \ \mathcal{F}'(- \Lambda) = 0 .  \label{eq3.4b}
\end{align}

\subsubsection{New solutions  of the form $a(t) = (\alpha \ e^{\lambda t} + \beta \ e^{-\lambda t})^\gamma$, \ ($k=0$)}
\label{3.2.2}

We are now going  to find some new cosmological solutions of the form
\begin{equation}
  a(t) = (\alpha e^{\lambda t} + \beta e^{-\lambda t})^\gamma,  \quad \gamma \in \mathbb{R},  \label{eq3.7}
\end{equation}
that satisfy equation
\begin{equation}
  \Box \sqrt{R-2\Lambda} = q \sqrt{R-2\Lambda} , \qquad   q = \zeta \Lambda \neq 0. \label{eq3.8}
\end{equation}
Note that the corresponding Hubble parameter and the Ricci scalar are:
\begin{align*}
  &H(t) = \left( 1-\frac{2 \beta}{\alpha e^{2 \lambda  t}+ \beta }\right) \gamma \lambda, \\
  &R(t) = 6k(\alpha e^{\lambda t} + \beta e^{-\lambda t})^{-2 \gamma}+ \frac{12 \gamma \lambda^{2} \left(  \alpha^{2} \gamma e^{4 \lambda  t} -2 \alpha \beta (\gamma-1)e^{2 \lambda  t}+ \beta^{2}\gamma \right)}{\left(\alpha  e^{2 \lambda  t}+ \beta\right)^2}.
\end{align*}

The equality \eqref{eq3.8} can be expanded into equation
\begin{align}
& -9\left(\beta +\alpha  e^{2 \lambda  t}\right)^4 \left(A_0 + A_1 e^{2\lambda t} + A_2 e^{4 \lambda  t} \right) \nonumber \\
&+3 \left(\alpha  e^{\lambda  t}+\beta  e^{-\lambda t}\right)^{2
   \gamma } \left(\beta +\alpha  e^{2 \lambda  t}\right)^2 \Big(B_0 + B_1 e^{2\lambda t} + B_2e^{4 \lambda  t} +B_3e^{6 \lambda  t} + B_4 e^{8 \lambda  t} \Big) \nonumber \\
 &- \left(\alpha  e^{\lambda  t}+\beta  e^{-\lambda t}\right)^{4
   \gamma } \Big(C_0 + C_1 e^{2\lambda t} + C_2e^{4 \lambda  t} +C_3e^{6 \lambda  t} + C_4 e^{8 \lambda  t} + C_5 e^{10 \lambda  t}+ C_6 e^{12 \lambda  t}\Big)=0, \label{eq3.9}
\end{align}
where
\begin{equation} \label{eq3.10}
\begin{aligned}
A_0 &=k^{2}\beta^2 \left(q-2 \gamma^2\lambda^2\right), \\
A_1 &=2 k^{2}\alpha  \beta  \left(2 (\gamma -1) \gamma  \lambda^2 + q \right), \\
A_2 &=k^{2} \alpha^2 \left(q-2 \gamma^2\lambda^2\right),
\end{aligned}
\end{equation}
\begin{equation} \label{eq3.11}
\begin{aligned}
B_0 &= -k\beta^4 \left(2q-\gamma^2\lambda^2\right)\left(6 \gamma^2 \lambda^2-\Lambda \right), \\
B_1 &= -4k \alpha  \beta^3 \left(3 \gamma  \left(2 \gamma^{3}+ 7 \gamma^2 -9 \gamma +2\right)
   \lambda^4 + \gamma \Lambda \lambda^{2}+ q \left(6 \gamma  \lambda^2 -2 \Lambda \right)\right), \\
B_2 &= 2k \alpha^2 \beta^2 \left(3 \gamma  \left(6 \gamma^{3} + 28 \gamma^2 -44 \gamma +16\right) \lambda^4+ \gamma(\gamma-4)\Lambda \lambda^{2}+  6 q \left(2\gamma^2 \lambda^2 -4 \gamma  \lambda^2+\Lambda \right)\right), \\
B_3 &=-4k \alpha^{3}  \beta \left(3 \gamma  \left(2 \gamma^{3}+ 7 \gamma^2 -9 \gamma +2\right)
   \lambda^4 + \gamma \Lambda \lambda^{2}+ q \left(6 \gamma  \lambda^2 -2 \Lambda \right)\right), \\
B_4 &= -k \alpha^4 \left(2q-\gamma^2\lambda^2\right)\left(6 \gamma^2 \lambda^2-\Lambda \right),
\end{aligned}
\end{equation}
and
\begin{equation} \label{eq3.12}
\begin{aligned}
C_0 &= q \beta^6 \left(6 \gamma^2 \lambda^2-\Lambda \right)^{2}, \\
C_1 &= -6 \alpha  \beta^5 \left(6 \gamma^2 \lambda^2-\Lambda \right) \left(2 \gamma  \left(-6 \gamma^2 +7 \gamma -2\right)
   \lambda^4 + q \left( 2 \gamma^{2}\lambda^2 -4 \gamma  \lambda^2 + \Lambda \right)\right), \\
C_2 &=-3 \alpha^2 \beta^4 \Big(48 \gamma^{2}  \left(12 \gamma^{3} -20 \gamma^2 +9 \gamma -1\right) \lambda^6 +
16 \gamma(2\gamma-1)\Lambda \lambda^{4} \\
&+ 4 q \left(3\gamma^4 \lambda^4 -12 \gamma^{2}\lambda^4 \right)+ q \Lambda\left(-4\gamma^2 \lambda^2 +32 \gamma  \lambda^2 -5 \Lambda
\right)\Big), \\
C_3 &=4 \alpha^3 \beta^3 \Big( 108 \gamma^{2}  \left(6\gamma^{3} -11\gamma^2 +8 \gamma -2\right) \lambda^6 +
 6 \gamma(6 \gamma^{2}-15 \gamma+6)\Lambda \lambda^{4} \\
 &+ 36 q \left(\gamma^4 \lambda^4 -2\gamma^{3}\lambda^4 +2 \gamma^{2}\lambda^4\right)+ q \Lambda\left(12\gamma^2 \lambda^2 -36 \gamma  \lambda^2 +5 \Lambda \right)\Big), \\
C_4 &=-3 \alpha^4 \beta^2 \Big(48 \gamma^{2}  \left(12 \gamma^{3} -20 \gamma^2 +9 \gamma -1\right) \lambda^6 +
16 \gamma(2\gamma-1)\Lambda \lambda^{4}\\
&+ 4 q \left(3\gamma^4 \lambda^4 -12 \gamma^{2}\lambda^4 \right)+ q \Lambda\left(-4\gamma^2 \lambda^2 +32 \gamma  \lambda^2 -5 \Lambda \right)\Big), \\
C_5 &= -6 \alpha^{5}  \beta \left(6 \gamma^2 \lambda^2-\Lambda \right) \left(2 \gamma  \left(-6 \gamma^2 +7 \gamma -2\right)
   \lambda^4 + q \left( 2 \gamma^{2}\lambda^2 -4 \gamma  \lambda^2 + \Lambda \right)\right), \\
C_6 &= q \alpha^6 \left(6 \gamma^2 \lambda^2-\Lambda \right)^{2}.
\end{aligned}
\end{equation}
It is evident that equation \eqref{eq3.9} is satisfied if
\begin{align}
  A_0 &=A_1 = A_2 =0, \qquad B_0=B_1 = B_2= B_3 = B_4 =0 ,   \label{eq3.13} \\
  C_0 &=C_1=C_2= C_3=C_4=C_5=C_6=0.\label{eq3.14}
\end{align}

If $\alpha\beta\neq 0, $   $R\neq 2\Lambda$ and $q \neq 0$ then equations  \eqref{eq3.13} and \eqref{eq3.14} are satisfied in the
flat space when
\begin{align}
 \gamma=\frac{2}{3},\ \ q= \frac{3}{8} \Lambda,\ \ \lambda = \pm \sqrt{\frac{3}{8} \Lambda} ,\ \ k = 0.  \label{eq3.15} \\
\end{align}

When $\alpha \neq 0$ and $\beta \neq 0$, we have the following  two special solutions:
\begin{align}
 &a_3 (t) = A\ \cosh^{\frac{2}{3}}{\big(\sqrt{\frac{3}{8} \Lambda}\ t\big)} , \quad  k= 0 , \quad \FF\big(\frac{3}{8} \Lambda\big)=-1, \ \FF'\big(\frac{3}{8} \Lambda\big)=0 , \label{eq3.17a} \\
 &a_4 (t) = A\ \sinh^{\frac{2}{3}}{\big(\sqrt{\frac{3}{8} \Lambda}\ t\big)} , \quad  k= 0 ,  \quad \FF\big(\frac{3}{8} \Lambda\big)=-1, \ \FF'\big(\frac{3}{8} \Lambda\big)=0 . \label{eq3.17b}
\end{align}
Solution similar to $a_3 (t)$ is also obtained  in \cite{kolar} nonlocal gravity  model of the form $R - 2\Lambda + R \mathcal{F} (\Box) R .$

\subsubsection{New solutions  of the form $a(t) = (\alpha \ \sin{\lambda t} + \beta \ \cos{\lambda t} )^\gamma$, \ ($k=0$)}

Note that $a(t) = \alpha \ \sin{\lambda t} + \beta \ \cos{\lambda t}$  can be presented as $a(t) = \bar{\alpha} \ e^{i\lambda t} + \bar{\beta} \
e^{-i\lambda t}$ with connections $\bar{\alpha} = \frac{\beta}{2} + \frac{\alpha}{2 i}, \ \  \bar{\beta} = \frac{\beta}{2} - \frac{\alpha}{2 i} ,$
i.e. $\bar{\alpha} = \bar{\beta}^* \ .$  As a consequence of this property, one can use results obtained in the previous subsubsection by replacement
$\lambda \to i \lambda$.

When $\alpha \neq 0$ and $\beta \neq 0$ there are two possibilities for $\gamma$: $\gamma = \frac{2}{3}$ and  $\gamma = \frac{1}{2}$.
Since $\alpha \ \sin{\lambda t} + \beta \ \cos{\lambda t}$ has negative as well as positive values, hence only $\gamma = \frac{2}{3}$ can be applied.
Taking $\beta = \alpha$ or $\beta = -\alpha$, and $\alpha^\frac{2}{3} = A$, we can write the following two  solutions:
\begin{align}
 &a_5 (t) = A\ \Big(1 + \sin{\big(\sqrt{-\frac{3}{2} \Lambda}\ t\big)} \Big)^{\frac{1}{3}}, \
           k= 0 , \quad \FF\big(\frac{3}{8} \Lambda\big)=-1, \ \FF'\big(\frac{3}{8} \Lambda\big)=0 ,  \label{eq3.18a} \\
 &a_6 (t) = A\ \Big(1 - \sin{\big(\sqrt{-\frac{3}{2} \Lambda}\ t\big)} \Big)^{\frac{1}{3}}, \
           k= 0 , \quad \FF\big(\frac{3}{8} \Lambda\big)=-1, \ \FF'\big(\frac{3}{8} \Lambda\big)=0 . \label{eq3.18b}
 \end{align}

When $\alpha = 0$ or $\beta = 0$, we have also two cosmological solutions with $\gamma = \frac{2}{3}$ and $k = 0$. These solutions are:
\begin{align}
 &a_{7} (t) = A\  \sin^{\frac{2}{3}}{\big(\sqrt{-\frac{3}{8} \Lambda}\ t\big)} , \quad   k= 0 , \quad \FF\big(\frac{3}{8} \Lambda\big)=-1, \ \FF'\big(\frac{3}{8} \Lambda\big)=0 ,  \label{eq3.19a} \\
 &a_{8} (t) = A\  \cos^{\frac{2}{3}}{\big(\sqrt{-\frac{3}{8} \Lambda}\ t\big)}   ,
      \quad     k= 0 , \quad \FF\big(\frac{3}{8} \Lambda\big)=-1, \ \FF'\big(\frac{3}{8} \Lambda\big)=0 . \label{eq3.19b}
\end{align}

Note that there are the following connections:
\begin{align}
a_5 (t) = \sqrt[3]{2}\ a_8 \Big(\frac{\pi}{4} \sqrt{-\frac{8}{3\Lambda}} - t \Big), \qquad a_6 (t) = \sqrt[3]{2}\ a_7 \Big(\frac{\pi}{4} \sqrt{-\frac{8}{3\Lambda}} - t \Big).  \label{eq3.19c}
\end{align}

\subsection{Cosmological solutions in the closed and open universe \ ($k =\pm 1$)}

We found three vacuum solutions in the closed and open FLRW space.

\subsubsection{Cosmological solution  of the form $a(t) = A\ e^{\pm \sqrt{\frac{\Lambda}{6}} t}$, \ ($k = \pm 1$)}

If in the above Section \ref{3.2.2}  $\alpha \neq 0, \beta =0$ or $\alpha =0, \beta \neq 0$ then there is a new solution which is the same as already obtained one,
$a_{9}(t) = A\ e^{\pm \sqrt{\frac{\Lambda}{6}} t} , \ \ \  k = \pm 1 . $

In the previous paper \cite{dimitrijevic10}, we presented the following exact solution:
\begin{align} \label{eq3.5}
a_{9}(t) = A\ e^{\pm \sqrt{\frac{\Lambda}{6}} t} , \quad k = \pm 1, \quad \mathcal{F}(\frac{1}{3} \Lambda) = -1 , \ \ \mathcal{F}'(\frac{1}{3} \Lambda) = 0,
\quad \Lambda > 0,
\end{align}
of the same EoM \eqref{eq2.12} with property \eqref{eq3.2}. In this case
\begin{align}  \label{eq3.6a}
&\Box = -\frac{\partial^2}{\partial t^2} - 3 H(t)\ \frac{\partial}{\partial t} , \quad H (t) = \pm \sqrt{\frac{\Lambda}{6}} , \\
&R(t) = \frac{6 k}{A^2}\ \exp{\Big(\mp \sqrt{\frac{2}{3}\Lambda}\ t\Big)} + 2\Lambda , \label{eq3.6b} \\
&\Box\ \sqrt{R - 2 \Lambda} = \frac{\Lambda}{3} \ \sqrt{R - 2 \Lambda} ,\quad \Box^{-1}\ \sqrt{R - 2 \Lambda} = \frac{3}{\Lambda}
\ \sqrt{R - 2 \Lambda} ,\label{eq3.6c} \\
&\FF\big(\frac{1}{3}\Lambda\big) = - 1 , \quad  \FF'\big(\frac{1}{3}\Lambda\big) = 0 .  \label{eq3.6c}
\end{align}

\subsubsection{New solutions  of the form $a(t) = (\alpha \ e^{\lambda t} + \beta \ e^{-\lambda t})^\gamma$, \ ($k=\pm 1$)}

Here we have the same general consideration until Eqs. \eqref{eq3.13} and \eqref{eq3.14} . When $\alpha \neq 0, \  \beta \neq 0 , \ R\neq 2\Lambda, \ q \neq 0$ and $k \neq 0$  we have
that Eqs. \eqref{eq3.13} and \eqref{eq3.14}  are satisfied if
\begin{equation}
 \gamma=\frac{1}{2},\ \ q=\frac{1}{3} \Lambda, \ \ \lambda= \pm \sqrt{\frac{2}{3} \Lambda} ,\ \ k \neq 0 .  \label{eq3....}
\end{equation}
The corresponding cosmological solutions are:
\begin{align}
&a_{10} (t) = A\ \cosh^{\frac{1}{2}}{\big(\sqrt{\frac{2}{3} \Lambda}\ t\big)} , \quad  k= \pm 1 ,  \quad \FF\big(\frac{1}{3} \Lambda\big)=-1, \ \FF'\big(\frac{1}{3} \Lambda\big)=0 , \label{eq3.17c} \\
 &a_{11} (t) = A\ \sinh^{\frac{1}{2}}{\big(\sqrt{\frac{2}{3} \Lambda}\ t\big)} , \quad  k= \pm 1 , \quad \FF\big(\frac{1}{3} \Lambda\big)=-1, \ \FF'\big(\frac{1}{3} \Lambda\big)=0 .  \label{eq3.17d}
\end{align}

\section{Discussion}
\label{Sec.4}

Here, we discuss cosmological solutions presented in the previous section.
We have practically $11$ exact vacuum background solutions of our nonlocal gravity model \eqref{eq2.1}.
These solutions can be divided into two classes: 1) $a_1 (t),\ a_2 (t),\ a_3 (t),\ a_4 (t),\ a_5 (t),\ a_6 (t),$ $a_{7} (t),\ a_{8} (t) $
which are solutions in the flat universe $(k= 0)$, and 2) $a_9 (t),\ a_{10} (t),\ a_{11} (t)$ that are the same solutions in both  closed $(k = + 1)$ and open space $(k = - 1)$.

It is useful to  introduce effective Friedmann equations to the above solutions:
\begin{equation}
\frac{\ddot{a}_i}{a_i} = - \frac{4\pi G}{3} (\bar{\rho}_i + 3 \bar{p}_i) + \frac{\Lambda_i}{3} \,,  \quad \frac{\dot{a}_i^2  + k}{a_i^2} = \frac{8\pi G}{3} \bar{\rho}_i
+ \frac{\Lambda_i}{3} , \quad i = 1, 2, ..., 11,  \label{eq4.01}
\end{equation}
where $\bar{\rho}_i$ and $\bar{p}_i$ are counterparts of the energy density and pressure in the standard model of cosmology, respectively.
$\Lambda_i$ is an effective cosmological constant, which differ from $\Lambda$ in $\Lambda$CDM cosmological model.  From  \eqref{eq4.01}
we have
\begin{equation}
\bar{\rho}_i(t) = \frac{3}{8\pi G} \Big(\frac{\dot{a}_i^2 + k}{a_i^2} - \frac{\Lambda_i}{3}\Big) , \quad \bar{p}_i (t) = \frac{1}{8\pi G}
\Big(\Lambda_i - 2\frac{\ddot{a}_i}{a_i} - \frac{\dot{a}_i^2 + k}{a_i^2}\Big) . \label{eq4.02}
\end{equation}
Then the  equation of  state is
\begin{equation}
\bar{p}_i (t) = \bar{w}_i(t) \, \bar{\rho}_i (t) , \label{eq4.03}
\end{equation}
where $\bar{w}_i(t)$ is the corresponding  effective state parameter.

We also use the following Planck-2018 results for the
$\Lambda$CDM parameters \cite{Planck2018}:
  \begin{align}
   &t_0  = (13.801 \pm 0.024) \times 10 ^9 \text{yr -- age of the universe}, \label{eq4.04a} \\
   &H (t_0) = (67.40 \pm 0.50) \text{ km/s/Mpc  -- Hubble parameter}, \label{eq4.04b} \\
  &\Omega_m = 0.315 \pm 0.007  \text{-- matter density parameter}, \label{eq4.04c} \\
  &\Omega_\Lambda = 0.685 - \Lambda \ \text{density parameter}, \label{eq4.04d}\\
  &w_0 = - 1.03 \pm 0.03  \text{--  ratio of pressure to energy density}.  \label{eq4.04e}
  \end{align}

\subsection{Solutions $a_1(t)  = A\ t^{\frac{2}{3}}\ e^{\frac{\Lambda_1}{14} t^2}$ \ and \ $a_2(t)  = A\ e^{\frac{\Lambda_2}{6} t^2} ,$ \ ($k= 0, \Lambda_{1,2} \neq 0$) }

 These two solutions, with $k=0, \Lambda_{1,2} \neq 0 $,  were found earlier and presented in \cite{dimitrijevic10}. We are  going to discuss them again and point out some new features.
 Both are even functions of cosmic time $t$. As we will see, especially significant is solution  $a_1(t)  = A\ t^{\frac{2}{3}}\ e^{\frac{\Lambda_1}{14} t^2} .$

\subsubsection{Cosmological solution $a_1(t)  = A\ t^{\frac{2}{3}}\ e^{\frac{\Lambda_1}{14} t^2}$}
\label{DM}

This solution is singular at cosmic time $t=0$. There are  two constraints on nonlocal operator: $\mathcal{F}\big(-\frac{3}{7} \Lambda_1 \big) = - 1, \ \mathcal{F}'\big(-\frac{3}{7} \Lambda_1 \big) = 0.$

Note that the  scale factor $a_1 (t)$ is a product of $t^{\frac{2}{3}}$ and $e^{\frac{\Lambda_1}{14} t^2}$, where $t^{\frac{2}{3}}$ is scale
factor of the matter-dominated universe in Einstein theory of gravity, while $e^{\frac{\Lambda_1}{14} t^2}$ is a new scale factor that   induces accelerated expansion of the universe for $\Lambda_1 > 0.$  This exponential acceleration is similar to the ordinary de Sitter case. The effective cosmological constant $\Lambda_1$ contains the dark energy. Since  $\Lambda_1$  is introduced in this model by its construction, a solution with   accelerated expansion of the universe is not a surprise. The surprise is emergence of the  effects which are usually assigned to the dark matter in the $\Lambda$CDM model. This nonlocal de Sitter nature of the dark matter is illustrated by analysis below.

The corresponding Hubble parameter,  acceleration and the Ricci scalar  are:
\begin{align} \label{eq4.1a}
&H_1(t) = \frac{\dot{a_1}}{a_1} = \frac{2}{3}\frac{1}{t}  + \frac{1}{7} \Lambda_1 t , \\
&\ddot{a}_1(t) = \Big(-\frac{2}{9} \frac{1}{t^2} + \frac{1}{3} \Lambda_1 + \frac{1}{49} \Lambda_1^2 t^2\Big)\ a_1(t) , \label{eq4.1b} \\
&R_1(t) = 6 \Big(\frac{\ddot{a}_1}{a_1} + \big(\frac{\dot{a}_1}{a_1}\big)^2\Big) = \frac{4}{3}\frac{1}{t^2} + \frac{22}{7} \Lambda_1 + \frac{12}{49} \Lambda_1^2 t^2 , \label{eq4.1c}
\end{align}
where subscript "$1$" means that these quantities are related to the solution $a_1 (t)$.
 The first terms expressed as function of time in all equations \eqref{eq4.1a}, \eqref{eq4.1b} and \eqref{eq4.1c} are just what is in the case of the matter-dominated universe in the Einstein theory of gravity. Since $t^{\frac{2}{3}}$ mimics effects of matter without matter at the cosmic scale, we see that in this case (dark) matter emerges in nonlocal gravity model \eqref{eq2.1} through vacuum solution $a_1(t)$.

 It is also worth mentioning   that at present cosmic time $t_0 = 13.801 \times 10^9$yr expression \eqref{eq4.1a} gives an interesting  connection between $H(t_0) = 67.40$ km/s/Mpc and  the effective cosmological constant $\Lambda_1$   \cite{dimitrijevic10}, i.e. the following equation holds:
\begin{align} \label{eq4.2a}
H_1(t_0) = \frac{2}{3}\frac{1}{t_0}  + \frac{1}{7} \Lambda_1 t_0 ,
\end{align}
where here we take $H_1(t_0) = H (t_0)$. Given $H (t_0)$ and $t_0$ we computed  $\Lambda_1 = 1.05 \times 10^{-35}$s$^{-2}$ that differs from $\Lambda$ in $\Lambda$CDM model, where $\Lambda = 3 H^2 (t_0)\ \Omega_\Lambda = 0.98 \times 10^{-35}$s$^{-2} .$ We also computed $\ddot{a}_1(t_0)/{a}_1(t_0)  = 2.7\times 10^{-36} s^{-2}$ and $R(t_0) = 4.5 \times 10^{-35} s^{-2}$. Consequently $R_1 (t_0) - 2 \Lambda_1 = 2.4 \times 10^{-35} s^{-2} . $

 Another aspect of this cosmological solution, that is useful to discuss, concerns effective energy density $\bar{\rho}_1(t)$ and pressure $\bar{p}_1(t)$.
 Replacing solution $a_1 (t)$ in \eqref{eq4.02} with $k=0$, we have
 \begin{align}
&\bar{\rho}_1 (t) = \frac{3}{8 \pi G} \Big(H_1^2(t) -\frac{\Lambda_1}{3}\Big) = \frac{3}{8 \pi G} \Big( \frac{4}{9} t^{-2} -\frac{1}{7} \Lambda_1 + \frac{1}{49}\Lambda_1^2 t^2 \Big) ,  \label{eq4.3a} \\
&\bar{p}_1(t) =  \frac{\Lambda_1}{56 \pi G} \big(1 - \frac{3}{7} \Lambda_1 t^2\big).   \label{eq4.3b}
\end{align}
The first term in $\bar{\rho}_1 (t)$ is proportional to  $\frac{4}{9t^2}$ and consequently behaviors in the same way as in the usual matter-dominated universe.
Computation of  \eqref{eq4.3a} in $t = t_0$ gives  $\bar{\rho}_1 (t_0) = 2.26 \times 10^{-30} \ \frac{g}{cm^3} .$
Note that energy density $\rho(t_0)$ in Einstein theory of gravity with $\Lambda$-term is
\begin{align}
\rho (t_0) = \frac{3}{8 \pi G} \Big(H^2(t_0) -\frac{\Lambda}{3}\Big) = 2.68 \times 10^{-30} \ \frac{g}{cm^3} . \label{eq4.4}
\end{align}
Then we have
\begin{align}
\rho(t_0) - \bar{\rho}_1 (t_0) = \frac{\Lambda_1 -\Lambda}{8 \pi G} = \rho_{\Lambda_1} - \rho_\Lambda = 0.42 \times 10^{-30} \ \frac{g}{cm^3} , \label{eq4.5}
\end{align}
 where
 \begin{align}
 \rho_{\Lambda_1} = \frac{\Lambda_1}{8 \pi G} = 6.25 \times 10^{-30} \ \frac{g}{cm^3} , \quad
 \rho_{\Lambda} = \frac{\Lambda}{8 \pi G} = 5.83    \times 10^{-30} \ \frac{g}{cm^3}  \label{eq4.6}
\end{align}
are effective vacuum energy density of background solution $a_1 (t)$ and $\Lambda$CDM model, respectively.

Recall that the critical energy density $\rho_c$ is
\begin{equation}
\rho_c = \frac{3}{8\pi G} H^2(t_0) = 8.51\times 10^{-30} \ \frac{g}{cm^3}.  \label{eq4.7}
\end{equation}
and consequently
\begin{align}
 &\Omega_{\Lambda_1} = \frac{\rho_{\Lambda_1}}{\rho_c} = 0.734 , \quad \Omega_{\Lambda} = \frac{\rho_{\Lambda}}{\rho_c} = 0.685 , \quad
 \Delta\Omega_\Lambda = \Omega_{\Lambda_1} - \Omega_{\Lambda} = 0.049 \label{eq4.8a} \\
 &\Omega_m = \frac{\rho(t_0)}{\rho_c} = 0.315 , \quad \Omega_{m_1} = \frac{\bar{\rho_1}(t_0)}{\rho_c} = 0.266 , \quad \Delta\Omega_m = \Omega_m - \Omega_{m_1}
 = 0.049 . \label{eq4.8b}
\end{align}
According to \eqref{eq4.8a} and  \eqref{eq4.8b}, we obtain that $\Omega_{m_1} = 26,6 \% $  corresponds to dark matter and
$\Delta\Omega_m = \Delta\Omega_\Lambda = 4.9 \%$ is related to visible matter, what is in a very good agreement with the standard model of cosmology \cite{Planck2018}. One can also conclude that the effective cosmological constant $\Lambda_1$ contains the standard cosmological constant of the $\Lambda$CDM and something else what is related to the standard visible matter.

It is worth to summarize the above analysis. We used the cosmological parameters listed in  \eqref{eq4.04a} -- \eqref{eq4.04d} to compute the effective  cosmological constant $\Lambda_1$ and the effective energy density $\bar{\rho}_1 (t_0)$ in this nonlocal de Sitter model. Comparing calculated $\Omega_{m_1} = \frac{\bar{\rho_1}(t_0)}{\rho_c}$ with dark matter density parameter in the $\Lambda$CDM model we  concluded that
$\bar{\rho}_1 (t_0)$ corresponds to the dark energy density. We also found that $\Omega_m -\Omega_{m_1} = \Omega_{\Lambda_1} - \Omega_\Lambda$ and is related to the visible matter density parameter.

Now we can discuss effective pressure \eqref{eq4.3b}. At the beginning, effective  pressure is positive, i.e. $\bar{p}_1(0) =  \frac{\Lambda_1}{56 \pi G}$, then decreases  with time and equals zero at $t = \sqrt{\frac{7}{3 \Lambda_1}} = 4.71 \times 10^{17} \ s = 14,917 \times 10^9 yr $.
According to \eqref{eq4.03}, we have parameter $\bar{w}_1(t) = \frac{\bar{p}_1 (t)}{\bar{\rho}_1 (t)}$ which has future behavior in agreement with standard model of cosmology, i.e.  $\bar{w}_1(t \to \infty) \to - 1.$

Note that the Hubble parameter \eqref{eq4.1a} has minimum at $t_{min} = 21.1 \times 10^9 yr$ and it is $H_1 (t_{min}) = 61.72 km/s/Mpc$. From \eqref{eq4.1b} follows that the universe experiences decelerated expansion during matter dominance, that turns to acceleration at  time $t_{acc} = 7.84 \times 10^9 yr$ when
$\ddot{a} = 0.$

\begin{figure}[tbp]
\centering 
\includegraphics[width=.45\textwidth]{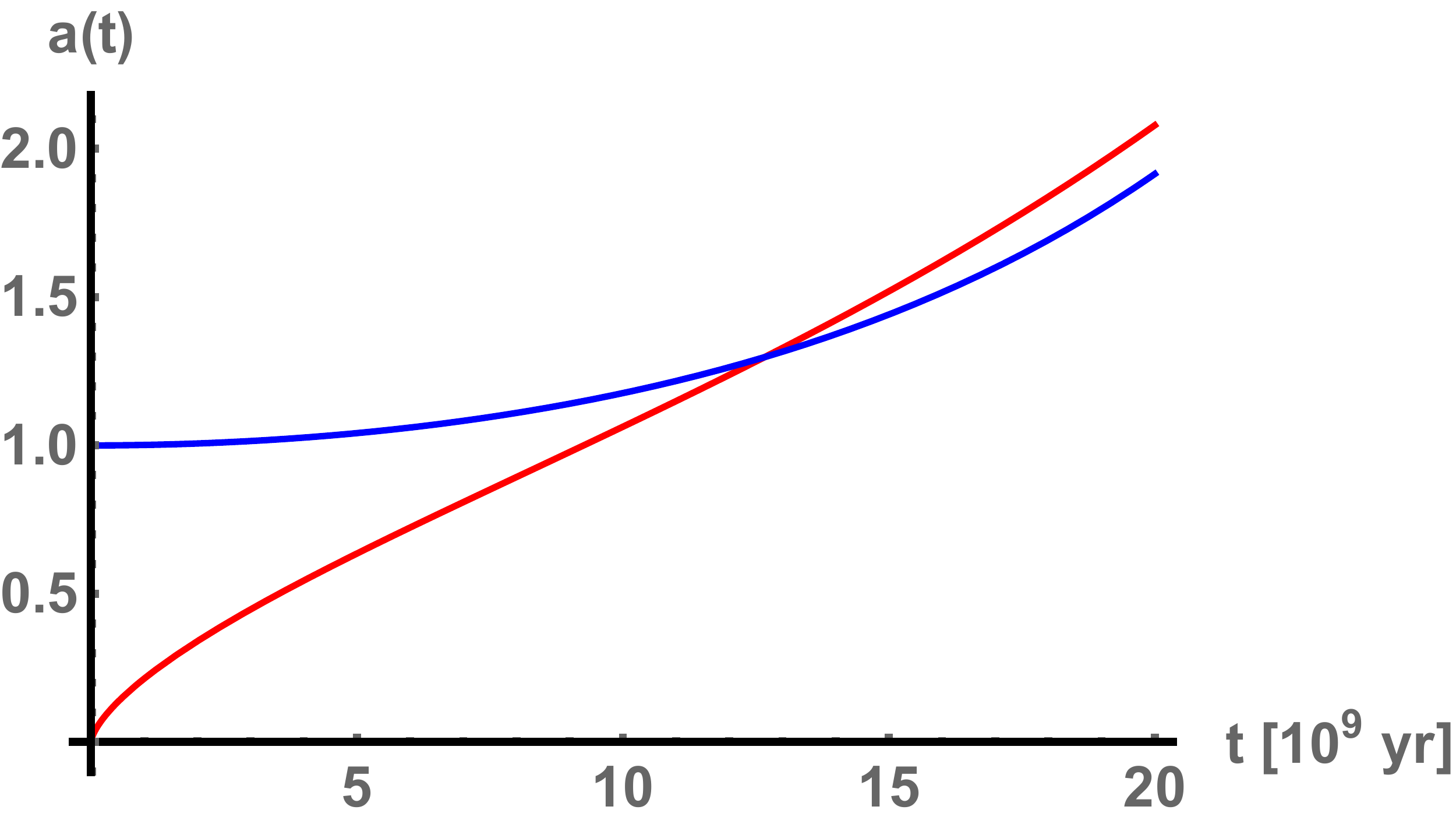}
\hfill
\includegraphics[width=.45\textwidth]{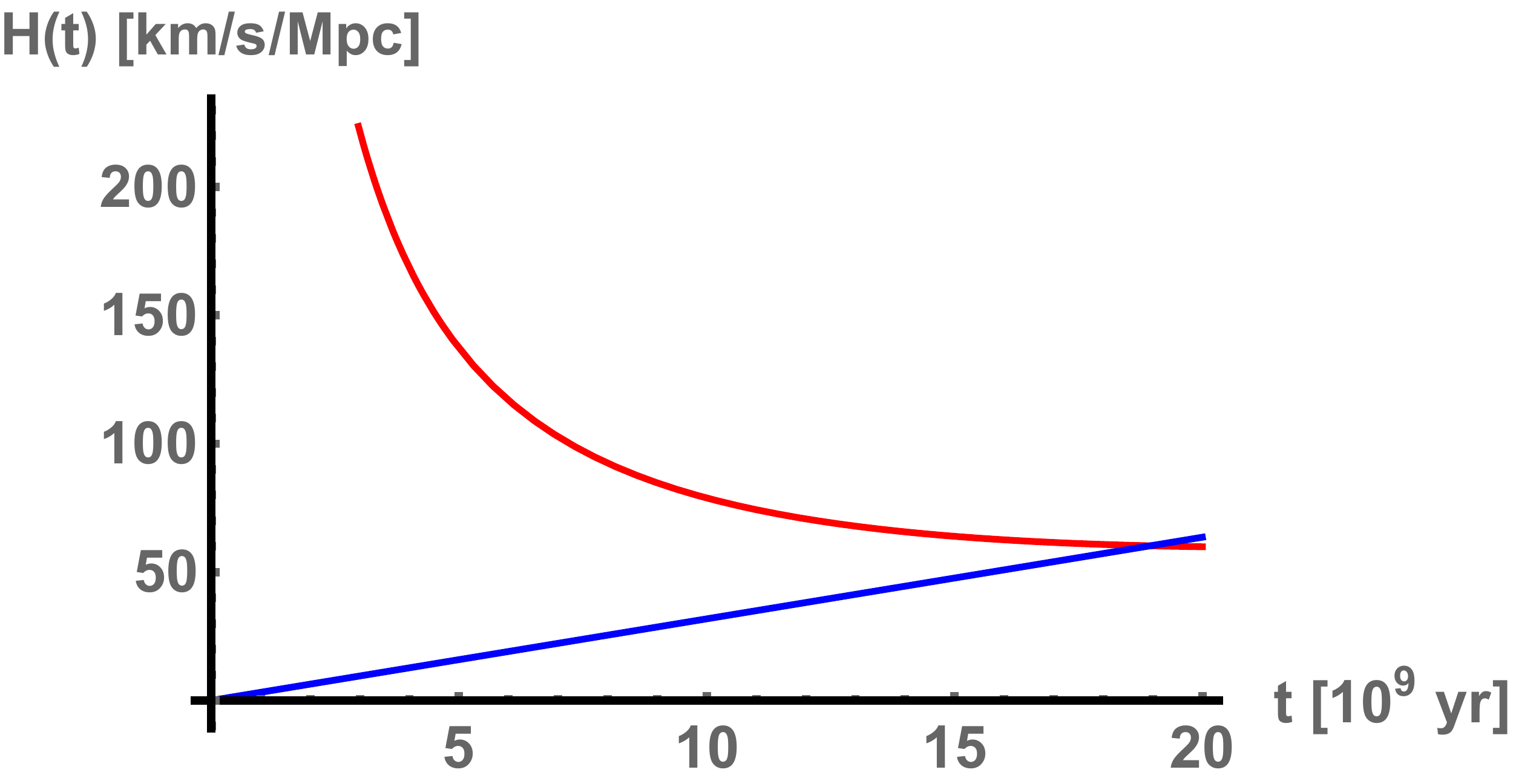}
\caption{\label{fig:i} Scale factors $a_1(t)  = A\ t^{\frac{2}{3}}\ e^{\frac{\Lambda_1}{14} t^2} $, (red) and $a_2(t)  = A\ e^{\frac{\Lambda}{6} t^2} $, (blue) are on the left side.  The corresponding Hubble parameters $H_1 (t)$ and $H_2 (t)$ are on the right side.}
\end{figure}


Presently, it  is not clear what role should play our other vacuum cosmological solutions  in the evolution of the universe. Hence, we cannot estimate effective cosmological constants $\Lambda_2, \Lambda_3, ..., \Lambda_{11}$ and we will use for them the standard cosmological constant $\Lambda$ in their graphic illustration and write almost everywhere  in the sequel $\Lambda$ instead of any  $\Lambda_2, \Lambda_3, ..., \Lambda_{11}$.

\subsubsection{Cosmological solution $a_2(t)  = A\ e^{\frac{\Lambda_2}{6} t^2}$}

As it is already said, we will use $\Lambda_2 = \Lambda$ and write $\Lambda$ instead of $\Lambda_2$.
This is  a nonsingular bounce solution  symmetric under  time change $t \to - t$, related to the flat space-time and similar to  $e^{\frac{\Lambda_1}{14} t^2}$  factor of  $a_1(t)$ cosmological solution. Equations of motion are satisfied when $\mathcal{F}(-\Lambda) = -1 , \ \ \mathcal{F}'(- \Lambda) = 0.$

The corresponding Hubble parameter is
\begin{equation}
H_2(t) = \frac{1}{3} \Lambda t,     \label{eq4.9}
\end{equation}
with $H_2 (0) = 0$ and depends linearly on the cosmic time $t$.    Equating $H_2 (t)$ with $H_1 (t)$ given by \eqref{eq4.1a}, and taking $\Lambda = 0.98 \times 10^{-35} s^{-2}$ from the $\Lambda$CDM model,  we  compute time $t = 5.98 \times 10^{17} s = 18.97 \times 10^9 yr$ when $H_2(t)$ reaches $H_1 (t)$ (see Fig. \ref{fig:i}).

The corresponding effective energy density and pressure are:
\begin{align}
 \bar{\rho}_{2}(t)=\frac{\Lambda  \left(\Lambda  t^2-3\right)}{24 \pi  G} , \quad
  \bar{p}_{2}(t)=\frac{\Lambda(1 -\Lambda  t^2)}{24 \pi  G} , \quad  \bar{w}_2 (t)_{t \to \infty} \to -1.
\end{align}
The equality $\bar{\rho}_{2}(t) = \bar{\rho}_{1}(t)$ gives the same value of the time as $H_2(t) = H_1 (t) ,$ that is $t = 18.97 \times 10^9 yr .$
When $\Lambda > 0$ then $\bar{\rho}_{2}(t) < 0$ for $t < \sqrt{\frac{3}{\Lambda}} $ and it is unlike that solution $a_2 (t)$ plays some cosmological role in this
cosmic  interval. However $a_2 (t)$ could be of some interest when $t > \sqrt{\frac{3}{\Lambda}}$ and $\Lambda >0$. Note that
$\bar{\rho}_{2}(t) > 0$  when $\Lambda < 0$ for any time $t$.

\subsection{Solutions  $a_3(t)  = A\ \cosh^{\frac{2}{3}}\big(\sqrt{\frac{3 \Lambda}{8}}\ t\big) $ and $a_4(t)  = A\ \sinh^{\frac{2}{3}}\big(\sqrt{\frac{3 \Lambda}{8}}\ t\big)$, ($k=0, \Lambda > 0 $)}

Solution $a_3 (t) = A\ \cosh^{\frac{2}{3}}{\big(\sqrt{\frac{3}{8} \Lambda}\ t\big)}$ satisfies equations of motion if
$\FF\big(\frac{3}{8} \Lambda\big)=-1, \ \FF'\big(\frac{3}{8} \Lambda\big)=0,$
The corresponding Hubble parameter, effective energy dense and pressure, are:
\begin{align}
 & H_{3}(t)=\sqrt{\frac{\Lambda}{6}} \ \tanh \big(\sqrt{\frac{3 \Lambda}{8}}\ t  \big), \\
 & \bar{\rho}_{3}(t)= \frac{\Lambda}{16 \pi  G}  \left(\tanh^2\big(\sqrt{\frac{3 \Lambda}{8}}\ t \big)-2\right), \quad
  \bar{p}_{3}(t)=\frac{\Lambda }{16 \pi  G}, \quad \bar{w}_3 (t)_{t \to \infty} \to -1.
\end{align}

In the similar way, solution $a_4(t)  = A\ \sinh^{\frac{2}{3}}\big(\sqrt{\frac{3 \Lambda}{8}}\ t\big)$ requires conditions
$\FF\big(\frac{3}{8} \Lambda\big)=-1, \ \FF'\big(\frac{3}{8} \Lambda \big) = 0$ to satisfy EoM.
The related Hubble parameter, and the analogs of energy density and pressure, are:
\begin{align}
 & H_{4}(t)= \sqrt{\frac{\Lambda}{6}} \ \coth \big(\sqrt{\frac{3 \Lambda}{8}}\ t  \big), \\
 & \bar{\rho}_{4}(t)= \frac{\Lambda}{16 \pi  G}  \left(\coth^2 \big(\sqrt{\frac{3 \Lambda}{8}}\ t  \big)-2\right), \quad
  \bar{p}_{4}(t)=\frac{\Lambda }{16 \pi  G} , \quad  \bar{w}_4 (t)_{t \to \infty} \to -1 .
\end{align}

\begin{figure}[tbp]
\centering 
\includegraphics[width=.45\textwidth]{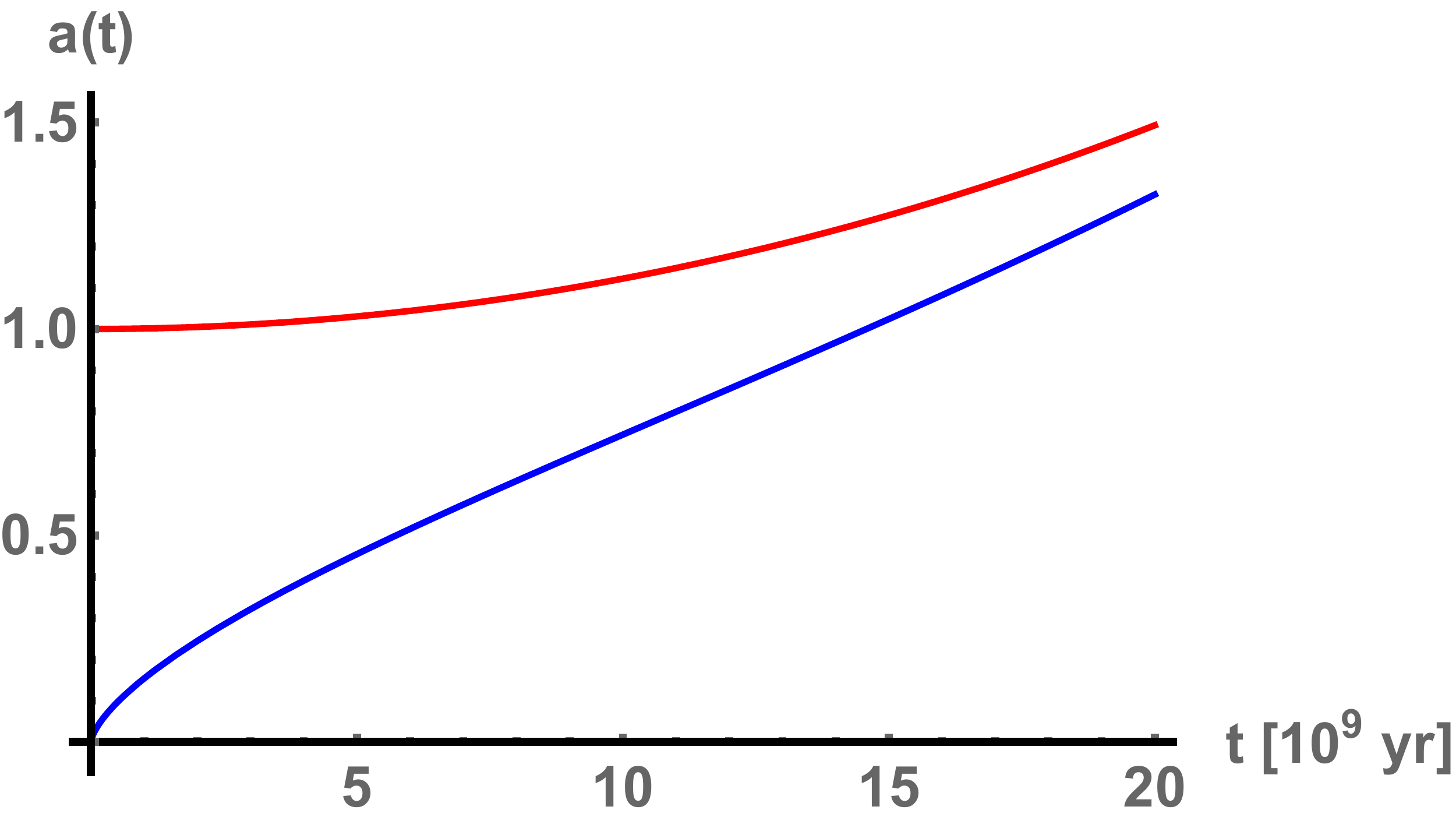}
\hfill
\includegraphics[width=.45\textwidth]{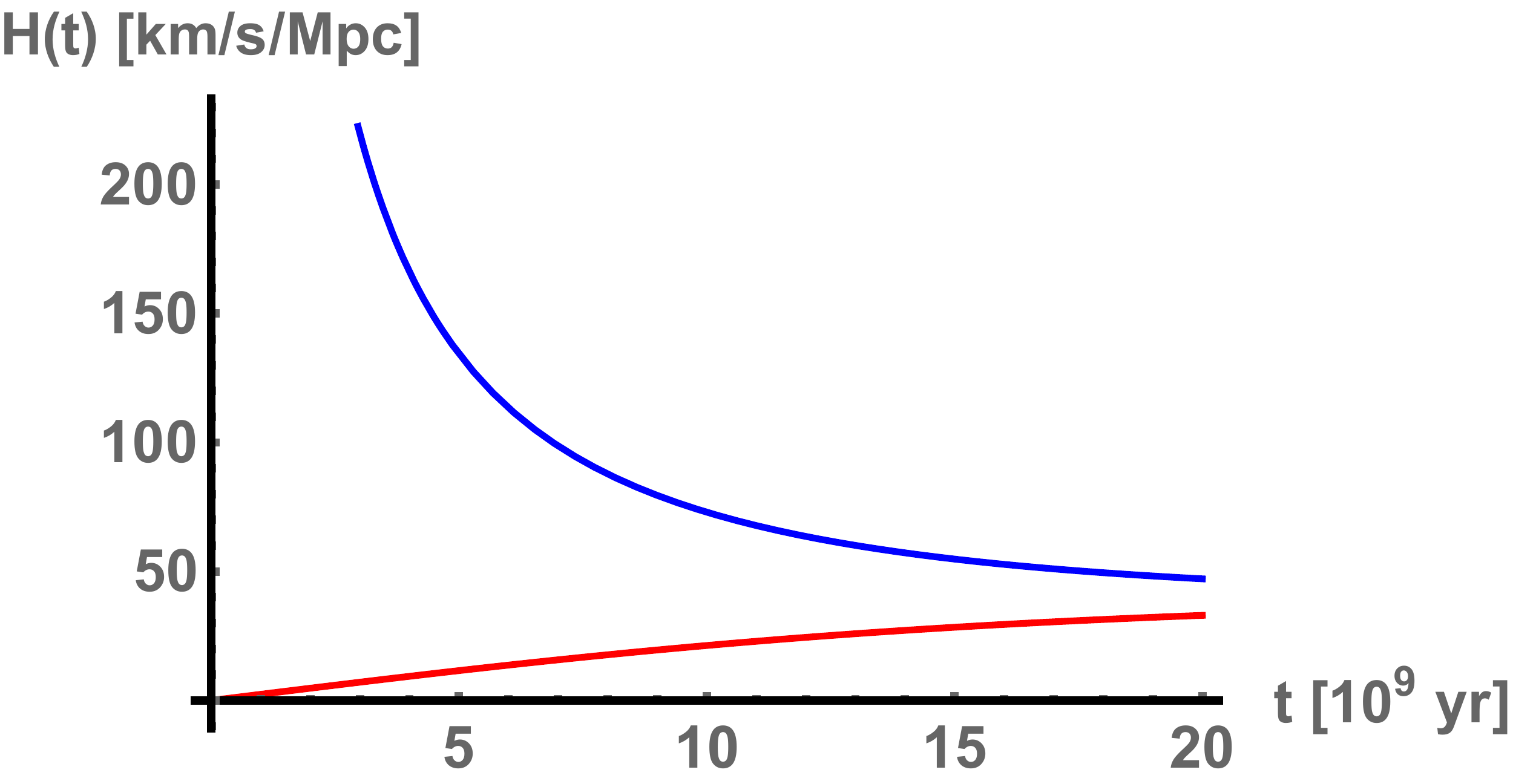}
\caption{\label{fig:2} Scale factors $a_3(t)  = A\ \cosh^{\frac{2}{3}}\big(\sqrt{\frac{3 \Lambda}{8}}\ t\big) $, (red) and  $a_4(t)  = A\ \sinh^{\frac{2}{3}}\big(\sqrt{\frac{3 \Lambda}{8}}\ t\big) $, (blue) are on the left side.  The corresponding Hubble parameters $H_3 (t)$ and $H_4 (t)$ are on the right side.}
\end{figure}

Both solutions $a_3 (t)$ and $a_4 (t)$ are even functions of cosmic time $t$. Solution $a_3 (t)$ is a nonsingular bounce solution, while $a_4 (t)$ is a singular one. For large cosmic times, these solutions become asymptotically close, see also Fig. \ref{fig:2}.

\subsection{Solutions $a_5(t)  = A\ \Big( 1 + \sin\big(\sqrt{-\frac{3 \Lambda}{2}}\ t\big)\Big)^{\frac{1}{3}} $  and $a_6(t)  = A\ \Big( 1 -\sin\big(\sqrt{-\frac{3 \Lambda}{2}}\ t\big)\Big)^{\frac{1}{3}}$, \ ($k=0, \Lambda < 0 $) }

Both  $a_5 (t)$ and $a_6 (t)$ are bounce solutions. They are also periodic solutions with periodicity  $T = 2\pi \sqrt{-\frac{2}{3 \Lambda}}$.
These two solutions could be interesting as vacuum backgrounds to study a toy cyclic cosmology. Both solutions satisfy equations of motion if
$\FF\big(\frac{3}{8} \Lambda\big)=-1$  and  $\FF'\big(\frac{3}{8} \Lambda\big)=0 .$ Solutions $a_5 (t)$ and $a_6 (t)$ replace each other under
change $t \to - t .$

\begin{figure}[tbp]
\centering 
\includegraphics[width=.45\textwidth]{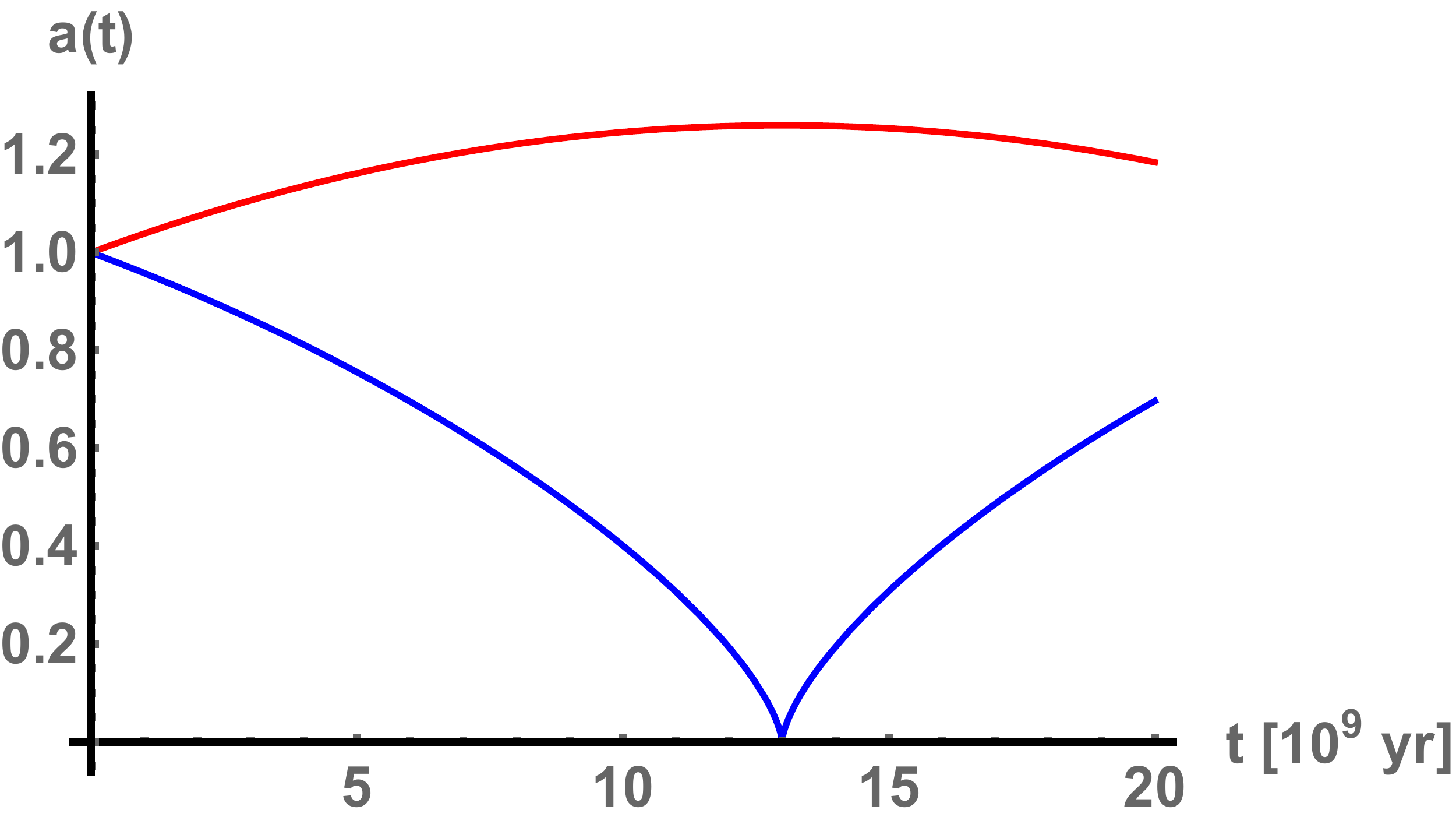}
\hfill
\includegraphics[width=.45\textwidth]{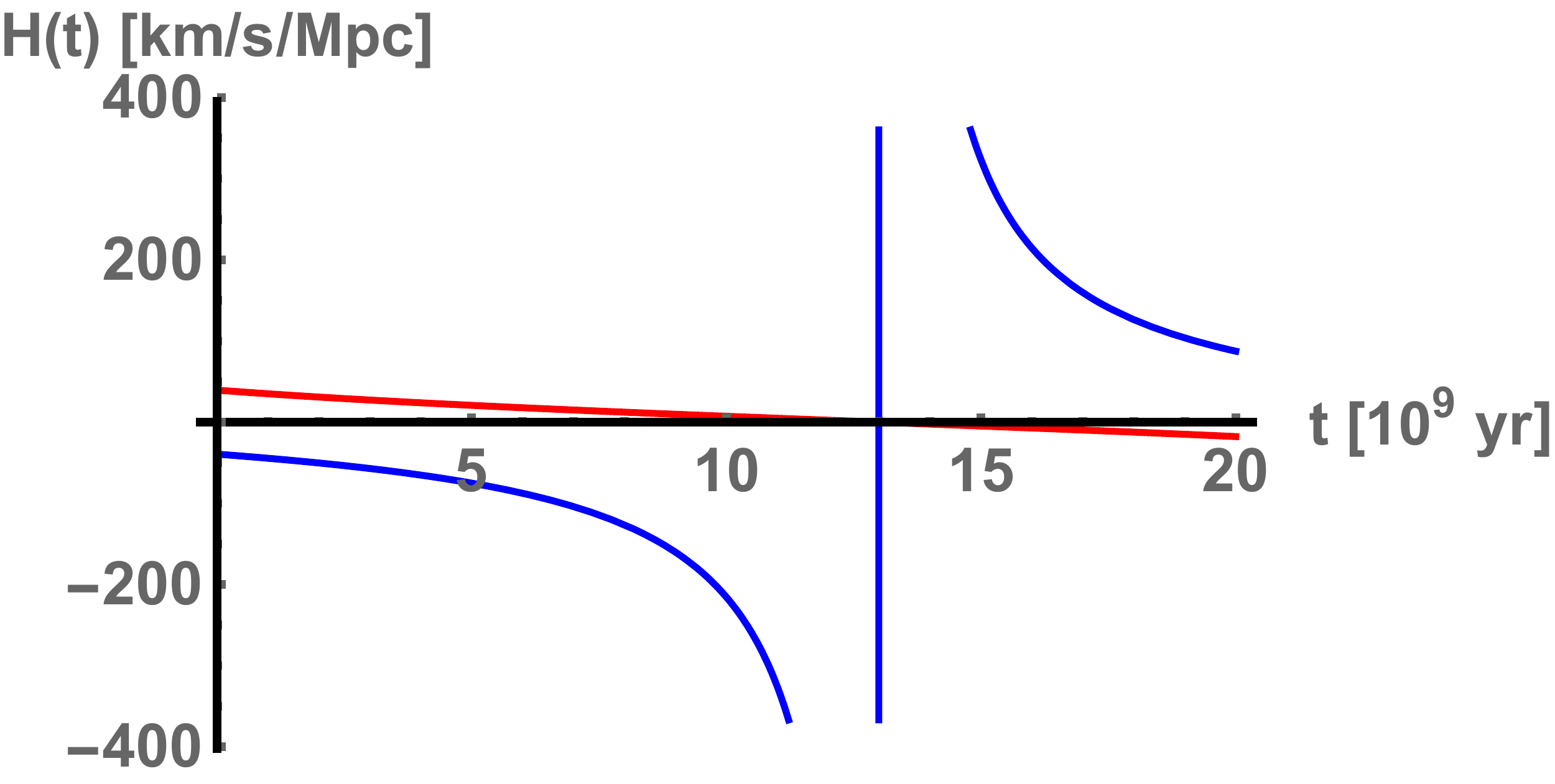}
\caption{\label{fig:3} Scale factors $a_5(t)  = A\ \Big( 1 + \sin\big(\sqrt{-\frac{3 \Lambda}{2}}\ t\big)\Big)^{\frac{1}{3}} $, (red) and $a_6(t)  = A\ \Big( 1 -\sin\big(\sqrt{-\frac{3 \Lambda}{2}}\ t\big)\Big)^{\frac{1}{3}} $, (blue) are on the left side.  The corresponding Hubble parameters $H_5 (t)$ and $H_6 (t)$ are on the right side. In these figures, we used $\Lambda = - 0.98 \times 10^{-35} s^{-2} .$}
\end{figure}

For solutions $a_5(t)  = A\ \Big( 1 + \sin\big(\sqrt{-\frac{3 \Lambda}{2}}\ t\big)\Big)^{\frac{1}{3}} $  and $a_6(t)  = A\ \Big( 1 -\sin\big(\sqrt{-\frac{3 \Lambda}{2}}\ t\big)\Big)^{\frac{1}{3}}$, we  have, respectively:
\begin{align}
 & H_{5}(t)= \sqrt{-\frac{\Lambda}{6}} \ \frac{\cos \left(\sqrt{-\frac{3\Lambda}{2}} \ t\right)}{\left(\sin \left(\sqrt{-\frac{3\Lambda}{2}} \ t\right)+1\right)}, \\
 & \bar{\rho}_{5}(t)= -\frac{\Lambda}{16 \pi  G}  \left(\frac{\cos^2\left(\sqrt{-\frac{3\Lambda}{2}} \ t\right)}{\left(\sin\left(\sqrt{-\frac{3\Lambda}{2}} \ t \right)+1\right)^2}+2\right),\qquad  \bar{p}_{5}(t)=\frac{\Lambda }{16 \pi  G},
\end{align}
\begin{align}
 & H_{6}(t)= \sqrt{-\frac{\Lambda}{6}} \   \frac{\cos \left(\sqrt{-\frac{3\Lambda}{2}} \ t\right)}{\left(\sin \left(\sqrt{-\frac{3\Lambda}{2}} \ t\right)-1\right)}, \\
 & \bar{\rho}_{6}(t)= -\frac{\Lambda}{16 \pi  G}  \left(\frac{\cos ^2\left(\sqrt{-\frac{3\Lambda}{2}} \ t\right)}{\left(\sin \left(\sqrt{-\frac{3\Lambda}{2}} \ t\right)-1\right)^2}+2\right), \qquad  \bar{p}_{6}(t)=\frac{\Lambda }{16 \pi  G}.
\end{align}
Solutions $a_5 (t)$ and $a_6(t)$, as well as related $H_5 (t)$ and $H_6 (t),$  are  illustrated in Fig. \ref{fig:3}.

\subsection{Solutions $a_{7}(t)  = A\ \sin^{\frac{2}{3}}\big(\sqrt{-\frac{3 \Lambda}{8}}\ t\big)$ and $a_{8}(t)  = A\ \cos^{\frac{2}{3}}\big(\sqrt{-\frac{3 \Lambda}{8}}\ t\big)$, \ ($k=0, \Lambda < 0 $) }

Solutions $a_7 (t)$  and $a_8 (t)$ are bounce ones with constraints on nonlocal operator $\FF\big(\frac{3}{8} \Lambda\big)=-1, \ \FF'\big(\frac{3}{8} \Lambda\big)=0$. Both solutions are even functions of $t$ and periodic with period
$T = 2 \pi \sqrt{-\frac{2}{3\Lambda}}$. These solutions can be interesting as backgrounds for toy cyclic universes in the flat space.

\begin{figure}[tbp]
\centering 
\includegraphics[width=.45\textwidth]{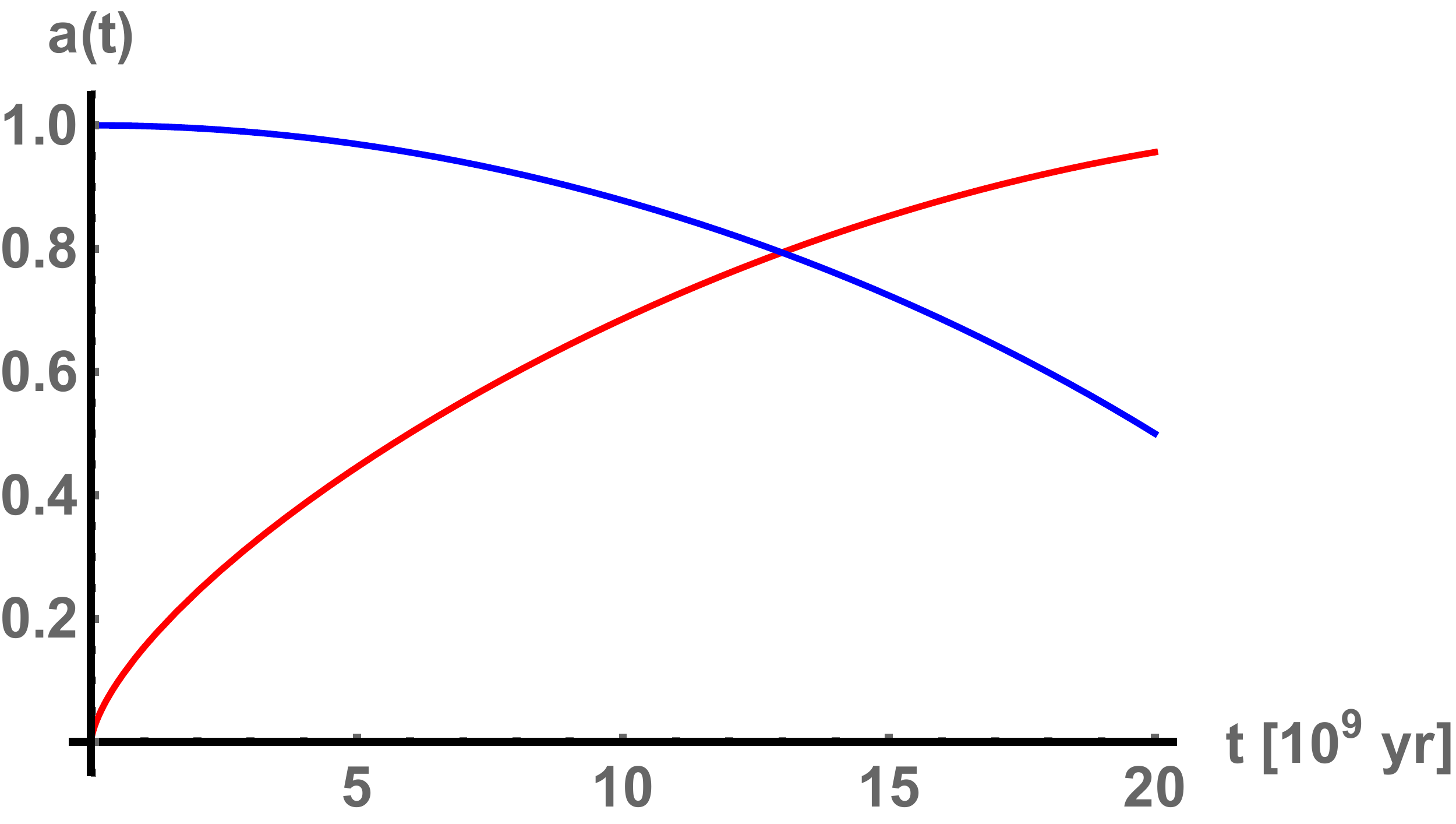}
\hfill
\includegraphics[width=.45\textwidth]{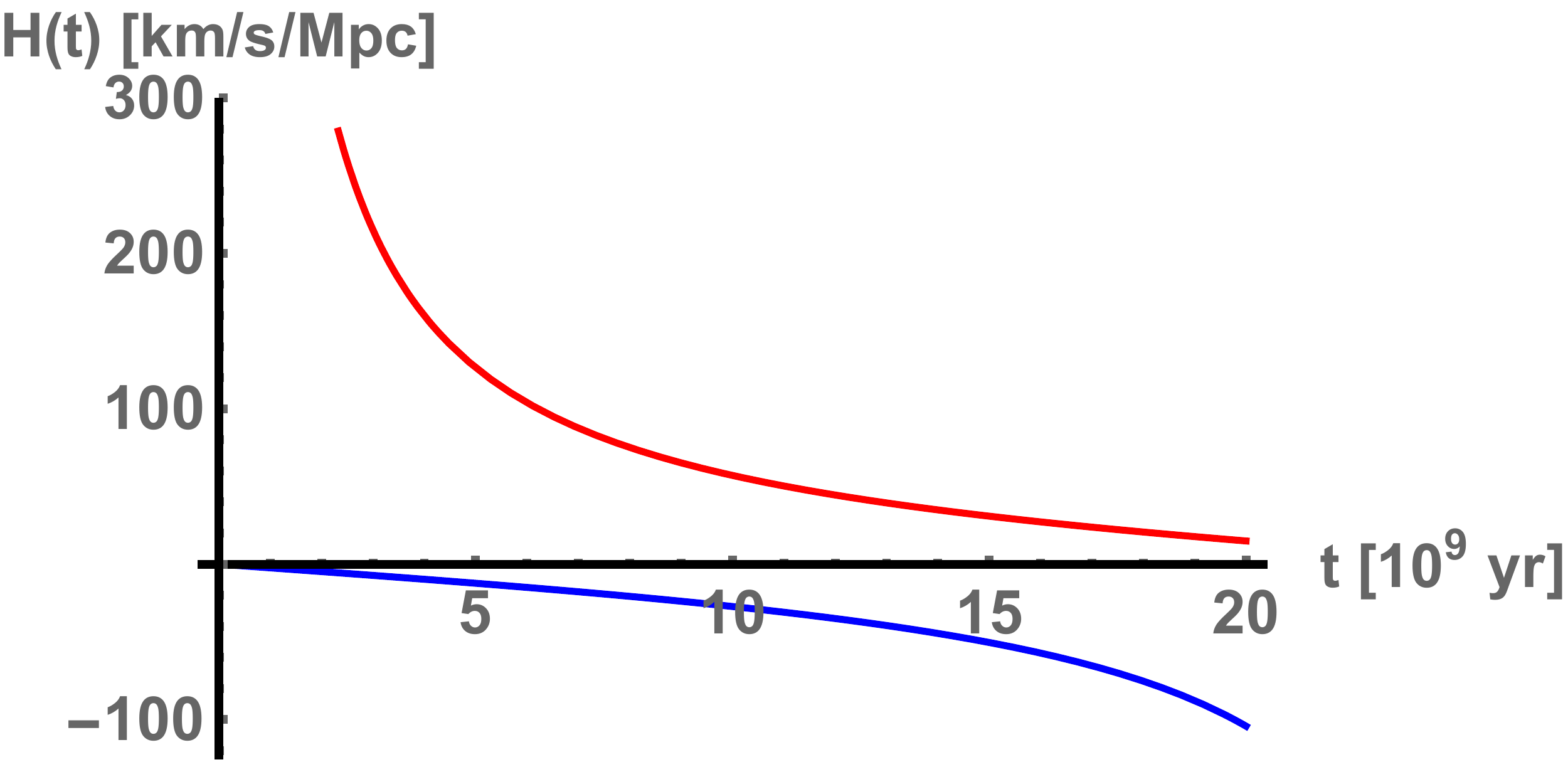}
\caption{\label{fig:4} Scale factors $a_{7}(t)  = A\ \sin^{\frac{2}{3}}\big(\sqrt{-\frac{3 \Lambda}{8}}\ t\big) $, (red) and $a_{8}(t)  = A\ \cos^{\frac{2}{3}}\big(\sqrt{-\frac{3 \Lambda}{8}}\ t\big) $, (blue) are on the left side.  The corresponding Hubble parameters $H_7 (t)$ and $H_8 (t)$ are on the right side.}
\end{figure}

The related Hubble parameters, energy density and pressure  are:
\begin{align}
 & H_{7}(t)= \sqrt{-\frac{\Lambda}{6}} \ \cot \left(\sqrt{-\frac{3 \Lambda}{8}}\ t\right), \\
 & \bar{\rho}_{7}(t)= -\frac{\Lambda}{16 \pi  G}  \left(\cot^2\left(\sqrt{-\frac{3 \Lambda}{8}}\ t\right)+2\right), \qquad
  \bar{p}_{7}(t)=\frac{\Lambda }{16 \pi  G}.
\end{align}
\begin{align}
 & H_{8}(t)= -\sqrt{-\frac{\Lambda}{6}} \ \tan \left(\sqrt{-\frac{3 \Lambda}{8}}\ t\right), \\
 & \bar{\rho}_{8}(t)= -\frac{\Lambda}{16 \pi  G}  \left(\tan^2\left(\sqrt{-\frac{3 \Lambda}{8}}\ t\right)+2\right), \qquad
 \bar{p}_{8}(t)=\frac{\Lambda }{16 \pi  G}.
\end{align}
See also Fig. \ref{fig:4}.

\subsection{Solutions  $a_9 (t) = A\ e^{\pm \sqrt{\frac{\Lambda}{6}} t}$, \ $a_{10}(t)  = A\ \cosh^{\frac{1}{2}}\big(\sqrt{\frac{3 \Lambda}{2}}\ t\big) $, and  $a_{11}(t)  = A\ \sinh^{\frac{1}{2}}\big(\sqrt{\frac{3 \Lambda}{2}}\ t\big)$, \ ($k=\pm 1, \Lambda > 0 $) }

First of all, note that all these three solutions are valid with the same form in both closed and open FLRW space. Solutions $a_9 (t)$ and
$a_{10} (t)$ are nonsingular bounce ones, while $a_{11} (t)$ is valid only for $t \geq 0$. All these solutions satisfy  equations of motion
under conditions $\mathcal{F}(\frac{1}{3} \Lambda) = -1$ and  $\mathcal{F}'(\frac{1}{3} \Lambda) = 0 .$

The corresponding Hubble parameter, energy density and pressure are as follows:
\begin{align}
 & H_{9}(t)=\pm \sqrt{\frac{\Lambda}{6}}, \\
 & \bar{\rho}_{9}(t)=\frac{6 k e^{\mp\sqrt{\frac{2}{3} \Lambda} \ t}-A^2 \Lambda }{16 \pi  A^2 G}, \qquad
  \bar{p}_{9}(t)=\frac{A^2 \Lambda -2 k e^{\mp\sqrt{\frac{2}{3} \Lambda} \ t}}{16 \pi  A^2 G} .
 \end{align}
\begin{align}
  H_{10}(t)&= \sqrt{\frac \Lambda6 } \tanh \left(\sqrt{\frac{2}{3}\Lambda}  t\right),\\
   \bar{\rho}_{10}(t)&= -\frac{A^2 \Lambda +A^2 \Lambda
   \cosh^{-2} \left(\sqrt{\frac{2}{3}\Lambda}
   t\right)-6 k \cosh^{-1}\left(\sqrt{\frac{2}{3}\Lambda} t\right)}{16 \pi  A^2 G},\\
   \bar{p}_{10}(t)&= -\frac{-3 A^2 \Lambda +A^2 \Lambda
   \cosh^{-2}\left(\sqrt{\frac{2}{3}\Lambda} t\right)+6 k \cosh^{-1} \left(\sqrt{\frac{2}{3}\Lambda} t \right)}{48 \pi  A^2 G},
\end{align}
\begin{align}
  H_{11}(t)&= \sqrt{\frac \Lambda6 } \coth \left(\sqrt{\frac{2}{3}\Lambda}  t\right),\\
   \bar{\rho}_{11}(t)&=\frac{-2 A^2 \Lambda +A^2 \Lambda  \coth
   ^2\left(\sqrt{\frac{2}{3}\Lambda} t\right)+6 k
   \sinh^{-1}\left(\sqrt{\frac{2}{3}\Lambda } t\right)}{16 \pi  A^2 G}, \\
   \bar{p}_{11}(t)&= \frac{2 A^2 \Lambda +A^2 \Lambda  \coth ^2\left(\sqrt{\frac{2}{3}}
   \sqrt{\Lambda } t\right)-6 k
   \sinh^{-1}\left(\sqrt{\frac{2}{3}} \sqrt{\Lambda }
   t\right)}{48 \pi  A^2 G} .
\end{align}

\begin{figure}[tbp]
\centering 
\includegraphics[width=.45\textwidth]{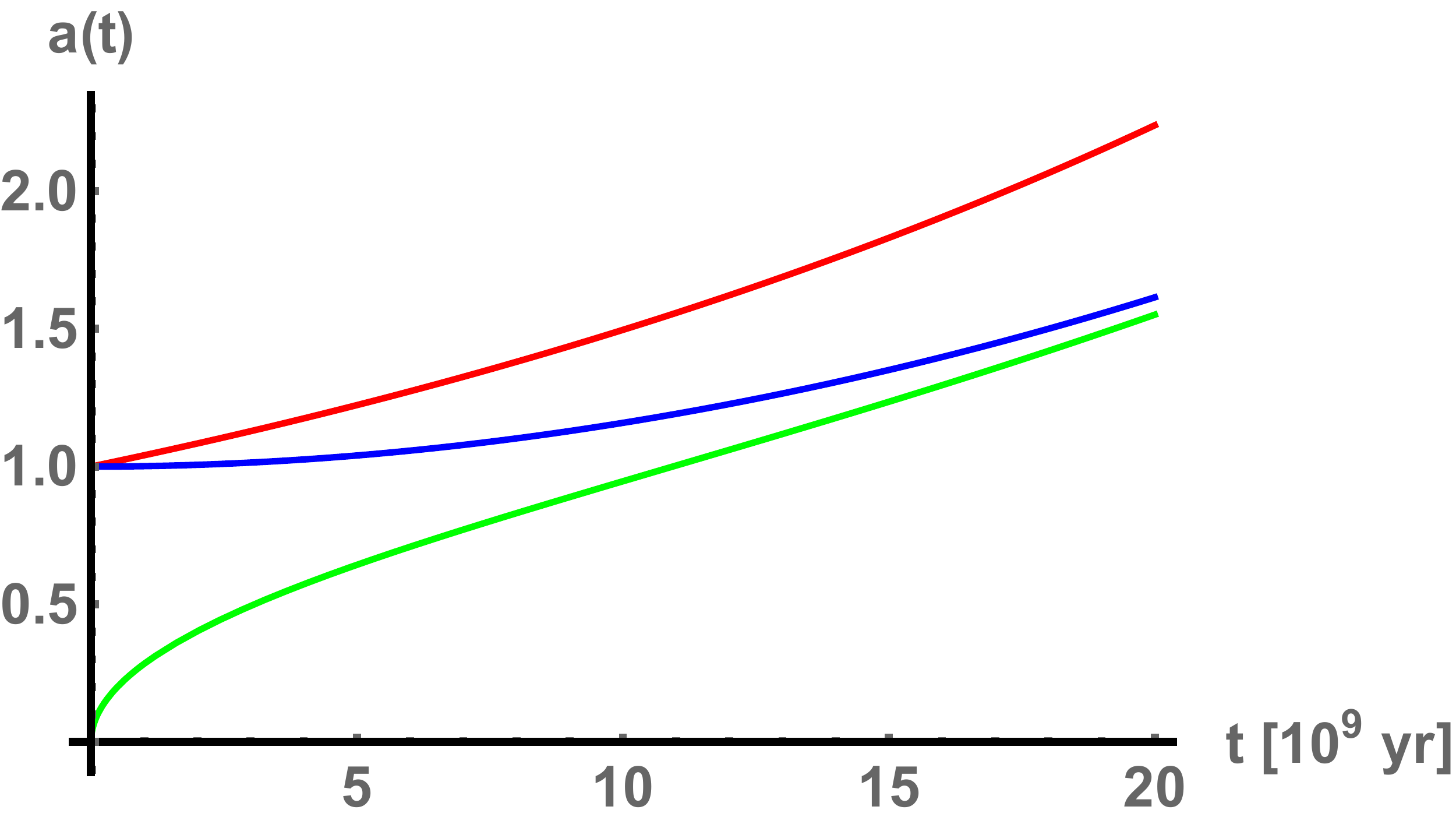}
\hfill
\includegraphics[width=.45\textwidth]{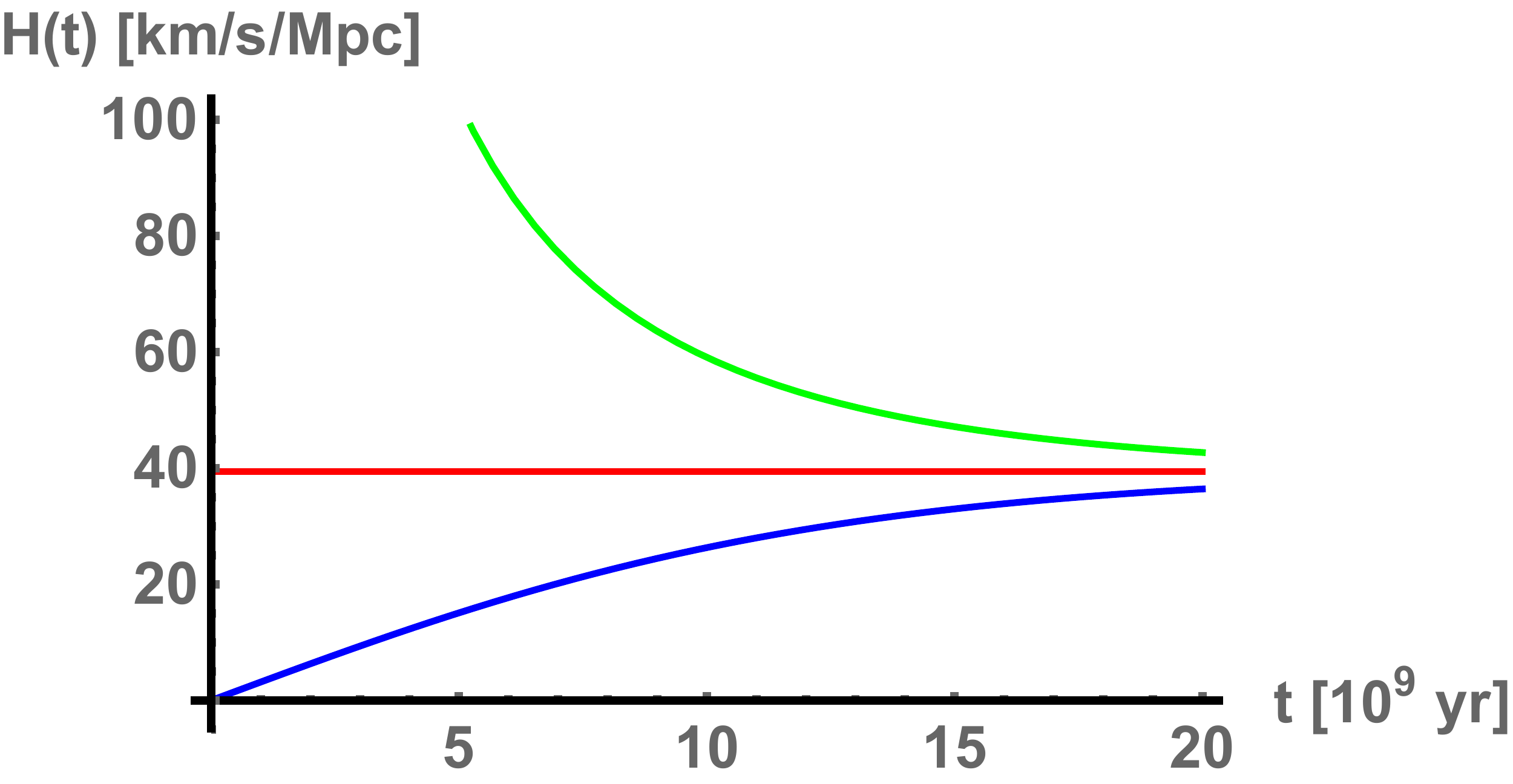}
\caption{\label{fig:5} Scale factors  $a_9 (t) = A\ e^{\pm \sqrt{\frac{\Lambda}{6}} t} $, (red) \ $a_{10}(t)  = A\ \cosh^{\frac{1}{2}}\big(\sqrt{\frac{2 \Lambda}{3}}\ t\big) $, (blue) and $a_{11}(t)  = A\ \sinh^{\frac{1}{2}}\big(\sqrt{\frac{2 \Lambda}{3}}\ t\big) $, (green) are on the left side.  The corresponding Hubble parameters $H_9 (t)$, $H_{10}(11)$ and $H_{11}(t)$ are on the right side.}
\end{figure}

The above solutions and the corresponding Hubble parameters are depicted in Fig. \ref{fig:5}.

\bigskip
\centerline{
-----------------------------------------------------
}

\bigskip

At the end of this section, it is worth noting  that we  have found no cosmological solutions of the form:
\begin{align*}
&a(t) = A\ t^n \ e^{\gamma t^m} , \quad k = +1, - 1 . \\
&a(t) = A\ t^n \ \cosh(\gamma t^m)   , \quad k = 0, +1, - 1 . \\
&a(t) = A\ t^n \ \sinh(\gamma t^m)   , \quad k =0, +1, - 1 .  \\
&a(t) = A\ t^n \ \sin(\gamma t^m)   , \quad k =0, +1, - 1 . \\
&a(t) = A\ t^n \ \cos(\gamma t^m)   , \quad k =0, +1, - 1 . \\
&a(t) = A\ t^n \tanh(\gamma t^m)    , \quad k =0, +1, - 1.
\end{align*}
where $m$ and $n$ are  rational numbers. Also, let us notice that there are cosmological investigation with nonlocal matter fields, see, e.g. \cite{arefeva2,koshelev2011,dragovich2}.

\section{Concluding Remarks}
\label{Sec.5}

We want to emphasize once again that this nonlocal de Sitter gravity model \eqref{eq2.1} maintains  $R -2\Lambda$ dependence like in the local case through  equality   $ R - 2\Lambda =\sqrt{R- 2\Lambda} \ \sqrt{R- 2\Lambda}$ and that nonlocalization  is introduced  replacing $\sqrt{R- 2\Lambda} \ \sqrt{R- 2\Lambda}$ by $\sqrt{R- 2\Lambda} {F} (\Box) \sqrt{R- 2\Lambda} .$ Dimensionless operator ${F} (\Box)$ is given by
expression \eqref{eq2.2} and in the more explicit form by Eq. \eqref{eq2.16} or \eqref{eq2.17}. In this nonlocal  de Sitter model, $\Lambda \neq 0$ is a parameter with dimensionality of the  cosmological constant and is an effective cosmological constat which may differ for different cosmological solutions.  There is not the Minkowski space solution in the general case of this model and it should not make problem in its use  to the Solar System, since $\Lambda$ has practically a very small value.

Another remarkable feature of this nonlocal de Sitter model consists  in the fact that it contains  $a_1(t)  = A\ t^{\frac{2}{3}}\ e^{\frac{\Lambda_1}{14} t^2}$ solution which  mimics effects related to dark matter and dark energy at the cosmological scale. Results presented in the section \ref{DM} are in  good agreement with parameters of the standard model of cosmology.
Dark matter emerges in this cosmological vacuum state as an effect of nonlocality in the presence of the corresponding effective cosmological constant $\Lambda_1$. Dark energy is related to the standard cosmological constant  $\Lambda$.

Main results presented in this article can be summarized as follows.
\begin{itemize}
\item Nonlocal de Sitter gravity model \eqref{eq1.2b}, introduced in \cite{dimitrijevic10} with operator $\mathcal{F} (\Box) = \sum_{n=1}^\infty f_n \Box^n ,$
is generalized by involving $\mathcal{F} (\Box) = \sum_{n=1}^\infty (f_n \Box^n + f_{-n} \Box^{-n}) .$ This generalized $\mathcal{F} (\Box)$
is presented in two explicit forms in \eqref{eq2.17}.
\item When $\Box \sqrt{R - 2\Lambda} = q  \sqrt{R - 2\Lambda}$ holds in \eqref{eq2.11a}, then equations of motion  have very simple form \eqref{eq2.12} which are satisfied with two conditions  \eqref{eq2.13} on nonlocal operator:
    $\mathcal{F} (q) = -1, \ \mathcal{F}' (q) = 0 .$
\item Vacuum solution $a_1(t)  = A\ t^{\frac{2}{3}}\ e^{\frac{\Lambda_1}{14} t^2}$ is investigated in details. Effective cosmological constant $\Lambda_1$ is computed using expression \eqref{eq4.2a} for the Hubble parameter. Then this $\Lambda_1$ is employed to calculate the effective energy density $\bar{\rho}_1 (t_0)$ according formula \eqref{eq4.3a}. Finally, we obtained  $\Omega_{m_1} = \frac{\bar{\rho_1}(t_0)}{\rho_c} = 26,6 \%$, which is counterpart of the dark matter density parameter, see \eqref{eq4.8b}.
\item Eight new exact  vacuum  cosmological solutions were found in Section \ref{Sec.3} by detail analysis of two classes of functions:
$a(t) = (\alpha \ e^{\lambda t} + \beta \ e^{-\lambda t})^\gamma$ and $a(t) = (\alpha \ \sin{\lambda t} + \beta \ \cos{\lambda t} )^\gamma .$
These solutions are discussed in Section \ref{Sec.4}.
\item All solutions are illustrated by  depicting the scale factors and Hubble parameters in figures. Also, effective energy density and pressure are presented for all vacuum solutions.

   \bigskip

We plan to continue investigation of this nonlocal de Sitter gravity model. Among the first tasks will be:
\begin{itemize}
\item investigation of cosmological solutions in the presence of  matter,
\item analysis of the solution around spherically symmetric body,
\item research towards  possible inflation aspects,
\item investigation of the  fluctuations,
\item further analysis of nonlocal operator $\mathcal{F} (\Box)$ from the point of view of general theoretical requirements, especially absence of possible ghosts.
\end{itemize}

\end{itemize}

At the end, it is worth noting that there is also cosmological research in application of nonlocal fields in the matter sector of general relativity, see, e.g. \cite{arefeva2,koshelev2011,dragovich2} and references therin.

\appendix
\section{Derivation of the equations of motion}

In this section we are going to present some main steps in derivation of the equations of motion for nonlocal gravity model given by action
\begin{equation} \label{action}
  S = \frac 1{16\pi G} \int \big(R-2\Lambda + P(R) \;\FF(\Box)\ Q(R)\big) \sqrt{-g}\; \dx,
\end{equation}
where $\FF(\Box) = \sum_{n=1}^{+\infty} f_n \ \Box^n + \sum_{n=1}^{+\infty} f_{-n} \ \Box^{-n}.$ The special case $P(R) = Q(R) = \sqrt{R - 2\Lambda}$
is nonlocal de Sitter gravity, that is under consideration in the previous sections. For some details in the case $\FF(\Box) = \sum_{n=1}^{+\infty} f_n \ \Box^n$ we refer to \cite{dimitrijevic9}.

To derive the expression for variation $\delta S$ with respect to the variation $\delta g^{\mu\nu}$, let us first recall  the variation of $\sqrt{-g}$,
the Christoffel symbol $\Gamma_{\mu\nu}^\lambda$, the Ricci tensor $R_{\mu\nu}$ and scalar curvature $R$:
  \begin{align}
  \delta \sqrt{-g} &= -\frac{1}{2} \sqrt{-g} \ g_{\mu\nu}\ \delta g^{\mu\nu} ,    \label{A.1} \\
  \delta \Gamma_{\mu\nu}^\lambda &= -\frac 12 \big( g_{\nu\alpha} \nabla_\mu \delta g^{\lambda \alpha} + g_{\mu\alpha} \nabla_\nu \delta g^{\lambda \alpha} - g_{\mu\alpha} g_{\nu\beta} \nabla^\lambda \delta g^{\alpha\beta}\big) , \label{A.2} \\
    \delta R_{\mu\nu} &= \nabla_\lambda \delta \Gamma_{\mu\nu}^\lambda - \nabla_\nu \delta \Gamma_{\mu\lambda}^\lambda,    \label{A.3} \\
    \delta R &= R_{\mu\nu} \delta g^{\mu\nu} - K_{\mu\nu} \delta g^{\mu\nu}, \quad K_{\mu\nu} = \nabla_\mu \nabla_\nu - g_{\mu\nu}\Box  .\label{A.4}
  \end{align}


Moreover, by application of the Stokes' theorem one obtains
that every scalar function $P(R)$ satisfies
    \begin{align}
        \int P K_{\mu\nu} \delta g^{\mu\nu} \sqrt{-g} \; \dx & = \int K_{\mu\nu} P \; \delta g^{\mu\nu} \sqrt{-g} \; \dx. \label{lem:hkdeltag0.3}
    \end{align}


From equation \eqref{A.4}  we get
\begin{equation}\begin{aligned}
\int P \delta R \sqrt{-g} \; \dx &= \int \left( R_{\mu\nu}P \delta g^{\mu\nu} - P K_{\mu\nu} \delta g^{\mu\nu}\right) \sqrt{-g} \; \dx \\
&= \int \left( R_{\mu\nu} P - K_{\mu\nu} P  \right)\delta g^{\mu\nu} \sqrt{-g} \; \dx.  \label{A.5}
\end{aligned}\end{equation}

Let us introduce the operator $\delta\Box$ by
\begin{equation}
  (\delta \Box) X  = \delta (\Box X) - \Box (\delta X).
\end{equation}
A straightforward calculation proves the following identity for all scalar functions $\HH$ and $\GG$.
    \begin{align}
        \int P (\delta \Box)Q  \sqrt{-g} \; \dx & =\frac 12 \int S_{\mu\nu} (P,Q)\; \delta g^{\mu\nu} \sqrt{-g} \; \dx, \label{lem:hkdeltag0.3}
    \end{align}
    where $S_{\mu\nu}(P,Q) = g_{\mu\nu} \nabla_\lambda P \nabla^\lambda Q + g_{\mu\nu}P \Box Q - 2\nabla_\mu P \nabla_\nu Q $.

On the other hand from $\delta (\Box \Box^{-1}) = 0$, we conclude that the variation of the inverse operator $\Box^{-1}$ is given by
\begin{equation}
  \begin{aligned}
     \delta \Box^{-1} &= - \Box^{-1} (\delta \Box) \Box^{-1}.
  \end{aligned}
\end{equation}

Moreover, operator $\Box^{-1}$ can be written as an formal integral $\Box^{-1} = \int_{0}^{+\infty} e^{-\alpha \Box} \mathrm d \alpha$.

The integral representation of $\Box^{-1}$ and repeated application of the Stoke's theorem yields
\begin{equation}
  \begin{aligned}
  \int P \Box^{-1} Q \sqrt{-g} \dx &= \int P \int_0^{+\infty} \sum_{n=0}^{+\infty} \frac{(-\alpha)^n}{n!} \Box^n Q  \sqrt{-g} \dx \;\mathrm d \alpha\\
    &= \int \int_0^{+\infty} \left(\sum_{n=0}^{+\infty} \frac{(-\alpha)^n}{n!} \Box^n P \right) Q  \sqrt{-g} \dx\; \mathrm d \alpha \\
    &= \int (\Box^{-1} P) Q  \sqrt{-g} \dx.
  \end{aligned}
\end{equation}

The variation $\delta S$ can be split into five parts
\begin{equation}
  \delta S = \frac 1{16\pi G} (I_0 + I_1 + I_2 + I_3 +I_4),
\end{equation}
where
\begin{equation}
\begin{aligned}
  I_0 = \int \delta (R-2\Lambda) \sqrt{-g}\; \dx,\\
  I_1 = \int \delta \HH\FF(\Box)\GG \sqrt{-g}\; \dx,\\
  I_2 = \int \HH\delta(\FF(\Box))\GG \sqrt{-g}\; \dx,\\
  I_3 = \int \HH\FF(\Box) \delta \GG \sqrt{-g}\; \dx,\\
  I_4 = \int \HH\FF(\Box)\GG \delta(\sqrt{-g})\; \dx.
\end{aligned}
\end{equation}
The first term $I_0$ is a well known
\begin{equation}
  I_0 = \int G_{\mu\nu} \delta g^{\mu\nu}\sqrt{-g}\; \dx.
\end{equation}
Integral $I_4$ is calculated directly and we get
\begin{equation}
  I_4 = -\frac 12 \int g_{\mu\nu} \HH\; \FF(\Box)\GG\; \delta g^{\mu\nu}\sqrt{-g}\; \dx.
\end{equation}
From equation \eqref{A.5}
integral $I_1$ gives,
\begin{equation}
\begin{aligned}
  I_1 &= \int \HHP\; \FF(\Box)\GG\; \delta R \sqrt{-g}\; \dx \\
  &= \int (R_{\mu\nu}-K_{\mu\nu})(\HHP\;\FF(\Box) \GG) \delta g^{\mu\nu}\sqrt{-g}\; \dx.
  \end{aligned}
\end{equation}
Integral $I_3$ after partial integration takes the same form as $I_1$ and therefore
\begin{equation}
\begin{aligned}
  I_3 &= \int \GGP\;\FF(\Box)\HH\; \delta R \sqrt{-g}\; \dx \\
  &= \int (R_{\mu\nu}-K_{\mu\nu})(\GGP\; \FF(\Box) \HH) \delta g^{\mu\nu}\sqrt{-g}\; \dx.
  \end{aligned}
\end{equation}
To calculate the last part $I_2$ we first introduce
 $J_n = \displaystyle \int \HH\;\delta(\Box^n)\GG\; \sqrt{-g}\; \dx$, then
\begin{equation}
  I_2 = \sum_{n= -\infty}^{+\infty} f_n J_n.
\end{equation}
Since $\Box^{0} = \mathrm{Id}$ and $\delta \mathrm{Id} = 0$ we conclude $J_0 = 0$.
We have to treat cases $n>0$ and $n<0$ separately.
Let $n\in \Natural$ then
\begin{equation}
  \begin{aligned}
    J_n &= \int \HH \; \delta(\Box^n)\GG\; \sqrt{-g}\; \dx \\
        &=\sum_{l=0}^{n-1} \int \Box^l \HH\; (\delta\Box) \Box^{n-1-l} \GG\; \sqrt{-g}\; \dx \\
        &=\frac 12 \sum_{l=0}^{n-1} \int S_{\mu\nu}(\Box^l \HH, \Box^{n-1-l} \GG) \delta g^{\mu\nu} \sqrt{-g}\; \dx.
  \end{aligned}
\end{equation}

To calculate $J_{-n}$ we do as follows
\begin{equation}
  \begin{aligned}
    J_{-n} &= \int \HH\;\delta(\Box^{-n})\GG \sqrt{-g}\; \dx \\
        &=\sum_{l=0}^{n-1} \int \Box^{-l} \HH\; (\delta\Box^{-1}) \Box^{-(n-1-l)} \GG\; \sqrt{-g}\; \dx \\
        &=-\sum_{l=0}^{n-1} \int \Box^{-(l+1)} \HH\; (\delta\Box) \Box^{-(n-l)} \GG\; \sqrt{-g}\; \dx \\
        &=-\sum_{l=0}^{n-1} \frac 12 \int S_{\mu\nu}(\Box^{-(l+1)} \HH, \Box^{-(n-l)} \GG) \delta g^{\mu\nu} \sqrt{-g}\; \dx.
  \end{aligned}
\end{equation}

Hence,
\begin{align}
  I_2 &= \frac 12\int \Omega_{\mu\nu} \delta g^{\mu\nu}\sqrt{-g}\; \dx, \\
  \Omega_{\mu\nu} &=  \sum_{n=1}^{+\infty} f_n \sum_{l=0}^{n-1} S_{\mu\nu}(\Box^l \HH, \Box^{n-1-l} \GG)  \nonumber \\ &-\sum_{n=1}^{+\infty} f_{-n} \sum_{l=0}^{n-1} S_{\mu\nu}(\Box^{-(l+1)} \HH, \Box^{-(n-l)} \GG).
\end{align}

Hence, we obtained
the variation of the action \eqref{action} in the following form
\begin{equation}\begin{aligned}
 \delta S&  = \frac 1{16\pi G} \int \hat G_{\mu\nu} \delta g^{\mu\nu} \sqrt{-g}\; \dx, \\
 \hat G_{\mu\nu} &= G_{\mu\nu} +\Lambda g_{\mu\nu} - \frac 12 g_{\mu\nu} \HH\; \FF(\Box)\GG + R_{\mu\nu}W -K_{\mu\nu}W + \frac 12 \Omega_{\mu\nu}, \\
 W &= 2 \HHP\;\FF(\Box)\GG, \\
  \Omega_{\mu\nu} &=  \sum_{n=1}^{+\infty} f_n \sum_{l=0}^{n-1} S_{\mu\nu}(\Box^l \HH, \Box^{n-1-l} \GG)  \\ &-\sum_{n=1}^{+\infty} f_{-n} \sum_{l=0}^{n-1} S_{\mu\nu}(\Box^{-(l+1)} \HH, \Box^{-(n-l)} \GG).
\end{aligned}
\end{equation}

Therefore equations of motion are given by
\begin{equation}\label{eom:box^-1}
\begin{aligned}
 G_{\mu\nu} +\Lambda g_{\mu\nu} -\frac 12 g_{\mu\nu} \HH \FF(\Box) \GG + R_{\mu\nu} W - K_{\mu\nu} W + \frac 12 \Omega_{\mu\nu} = 0.
\end{aligned}
\end{equation}

It is interesting to note that $ \nabla^\mu \hat G_{\mu\nu} =0$.

\acknowledgments
This research was partially funded by the Ministry of Education, Science and Technological Developments of the Republic of Serbia: grant number
451-03-68/2022-14/200104 with Faculty of Mathematics, and grant number 451-03-1/2022-14/4 with Teacher Education Faculty, University of Belgrade.














\end{document}